\documentclass[10pt,aps,pre,amsmath,amsfonts,amssymb,floatfix,superscriptaddress,notitlepage,nofootinbib]{revtex4}
\usepackage{mathrsfs} \usepackage{graphicx,color} 
\usepackage{multirow}
\usepackage{bm}

\font\msytw=msbm9 scaled\magstep1

\let\a=\alpha \let\b=\beta \let\g=\gamma \let\d=\delta
   \let\k=\kappa
\let\l=\lambda \let\m=\mu  \let\x=\xi 
\let\s=\sigma   \let\c=\chi
   
 \let\Th=\Theta  
   
\let\ee=\epsilon \let\r=\rho \let\th=\theta \let\io=\infty

\def\MM{{\cal M}} 
\def\CC{{\cal C}}\def\FF{{\cal F}} \def\HH{{\cal H}}
 
  \def\OO{{\cal O}}
\def\DD{{\cal D}} \def\SS{{\cal S}}

  \def\erf{\text{erf}}

\def\de{\mathrm d}
\newcommand{\bx}{\vec{x}}

\def\to{\rightarrow} \def\la{\left\langle} \def\ra{\right\rangle}

\def\RRR{\hbox{\msytw R}}

\newcommand{\beq}{\begin{equation}} \newcommand{\eeq}{\end{equation}}
\newcommand{\wh}{\widehat}

\begin{document}
\title{
Universality of the SAT-UNSAT (jamming) threshold \\
in non-convex continuous constraint satisfaction problems
}

\author{Silvio Franz}
\affiliation{LPTMS, Universit\'e Paris-Sud 11,
UMR 8626 CNRS, B\^at. 100, 91405 Orsay Cedex, France}

\author{Giorgio Parisi}
\affiliation{Dipartimento di Fisica,
Sapienza Universit\`a di Roma,
P.le A. Moro 2, I-00185 Roma, Italy}
\affiliation{
INFN, Sezione di Roma I, Nanotec -- CNR,
P.le A. Moro 2, I-00185 Roma, Italy
}

\author{Maksim Sevelev}
\affiliation{LPTMS, Universit\'e Paris-Sud 11,
UMR 8626 CNRS, B\^at. 100, 91405 Orsay Cedex, France}

\author{Pierfrancesco Urbani}
\affiliation{
Institut de physique th\'eorique, Universit\'e Paris Saclay, CNRS, CEA, F-91191 Gif-sur-Yvette, France
}

\author{Francesco Zamponi}
\affiliation{LPT,
\'Ecole Normale Sup\'erieure, UMR 8549 CNRS, 24 Rue Lhomond, 75005 Paris, France}

\begin{abstract}
Random constraint satisfaction problems (CSP) have been studied extensively using statistical physics techniques.
They provide a benchmark to study average case scenarios instead of the worst case one.
The interplay between statistical physics of disordered systems and computer science has brought new light into the realm of computational
complexity theory, by introducing the notion of {\it clustering} of solutions, related to replica symmetry breaking.
However, the class of problems in which clustering has been studied often involve 
discrete degrees of freedom: standard random CSPs are random $K$-SAT (\emph{aka} disordered Ising models) or random coloring problems (\emph{aka} disordered Potts models).
In this work we consider instead problems that involve continuous degrees of freedom.
The simplest prototype of these problems is the perceptron. 
Here we discuss in detail the full phase diagram of the model.
In the regions of parameter space where the problem is non-convex, leading to multiple disconnected clusters of solutions,
the solution is critical at the SAT/UNSAT threshold and lies in the same universality class of the jamming transition of soft spheres.
We show how the critical behavior at the satisfiability threshold emerges,
 and we compute the critical exponents associated to the approach to the transition from both the SAT and UNSAT phase.
 We conjecture that there is a large universality class of non-convex continuous CSPs whose SAT-UNSAT threshold is described 
 by the same scaling solution.
\end{abstract}

\maketitle

\section{Introduction}
The exact description of the glassy phases of high-dimensional sphere
systems
and the discovery that universal predictions at jamming match
finite dimensional observations~\cite{CKPUZ16} has renewed the interest in
the statistical physics of random constraint satisfaction problems (CSP)
with continuous variables. In CSPs, one
seeks assignments of a set of $N$ variables that satisfy a system of
constraints. Sphere systems clearly belong to this class: one needs to
find the positions of $N$ spheres inside a box with the conditions that the spheres
do not overlap. Particularly interesting in a statistical physics
perspective is the case where the constraints are taken at random from
some ensemble \cite{MZKST99,MPZ02,KMRSZ07}. In that case, quite generically in the limit of large
systems one observes a phase transition as the density of constraints
is increased passing from a satisfiable (SAT) phase where admissible configurations
exist to an unsatisfiable (UNSAT) phase, where the minimum number of unsatisfied
constraints is larger than zero. This SAT/UNSAT threshold is clearly analogue to
the jamming transition in soft-spheres~\cite{OSLN03,LNSW10}, that separates the low density
region where spheres do not overlap from the high density one where
no-overlap configurations do not exist.  The jamming transition in spheres has
highly universal features, with exponents that appear to be independent
of the dimension and of the protocol used to produce the jammed
configurations~\cite{OSLN03,LNSW10,Wy12,LDW13,CCPZ12,CCPZ15,CKPUZ16,GLS16,Ka16}. 
While some exponents, like e.g. the ones relating the
number of contacts or the pressure to the packing fraction in the UNSAT phase, take
simple semi-integer values~\cite{GLS16}, other exponents, e.g. the ones describing
the distributions of forces and the interparticle distances have non
trivial, presumably non-rational values~\cite{CKPUZ14}.  

In order to understand the
origin of this universality, it is important to study the SAT/UNSAT
transition in different CSP models. A crucial ingredient for jamming
is the continuous nature of variables. Jamming is the point where the
volume of the set of solutions to the problem continuously shrinks to
zero, and in its vicinity scaling laws can emerge. To this aim, in
\cite{FP15} it was suggested to study the random perceptron problem
\cite{DG88} as a prototype of a CSP with continuous variables, generalized to a region of non-convex
optimization. It was found that the nontrivial criticality and
universality of the SAT/UNSAT transition point was associated to
non-convexity. In the convex regime jamming is reached from a liquid,
ergodic phase and it is hypostatic and non-critical. In the non-convex
regime jamming is reached from a glassy phase, it is critical and in
the same universality class of spheres. This led to the conjecture that
it exists a large set of continuous CSP that belong to the same universality class.
The perceptron emerges therefore as the simplest continuous CSP where
glassy phenomena and jamming can be studied. This paradigm has been fruitfully applied to study the vibrational spectrum of glasses at low temperatures.
In \cite{FPUZ15} the spectrum of the Hessian matrix of the energy minima in the UNSAT phase of the perceptron has been computed
showing that it captures essential features of the vibrations of low temperature glasses. Furthermore, in \cite{AFP16} it has been shown how to study systematically the free energy landscape of the model using a Thouless-Anderson-Palmer approach \cite{TAP77} to obtain, in particular, the vibrational spectrum in the SAT phase. In~\cite{FS16}, the avalanches characterizing the glassy phase around jamming
have also been studied. Thanks to these studies, the non-convex perceptron now emerges as the simplest model that captures, at the mean field level, all the most important features of the glass
and jamming transitions.

The scope of this paper is to give a detailed account of the space of solutions
of the random perceptron model.
In particular we carefully study the scaling behavior close to jamming, coming both from the SAT and UNSAT phase.
The paper is organized as follows. 
In Sec.~\ref{sec:CCSP} we give a general formulation of continuous CSPs, we discuss the properties of the SAT-UNSAT transition,
and we briefly discuss the case of sphere packings to motivate some denominations that are used throughout the paper.
In Sec.~\ref{sec:perc_def} we define the random perceptron model, we introduce the replica method to solve it, and we give the main
equations needed for its study.
In Sec.~\ref{sec:PD} we discuss the zero-temperature phase diagram of the model, and
in Sec.~\ref{sec:SATUNSAT} we completely characterize the SAT/UNSAT transition (jamming) line and its critical properties.
Finally, we present concluding remarks and perspectives for future work.

\section{Continuous constraint satisfaction problems}
\label{sec:CCSP}

\subsection{Thermodynamic free energy, and the space of solutions}

In an  abstract form continuous CSPs can be formulated in the following
way: 
find a $N$-dimensional vector $\vec X =
\{X_i\}_{i=1\cdots N}\in \RRR^N$ that satisfies the set of $M$ constraints 
\begin{eqnarray}
  \label{eq:16}
  h_\mu(\vec X)>0 \ ,
  \qquad
  \mu=1,...,M \ .
\end{eqnarray}
The constraints are specified by some real 
functions $h_\mu(\vec X): \RRR^N\to \RRR$, which can be either deterministic or random 
(i.e. they may contain some quenched disorder).
One can
associate this problem to an optimization one,  defining a
Hamiltonian function taking positive values if at least one constraint
is not satified and zero if all the constraints are satisfied. There
is a large choice for such a Hamiltonian; here in analogy with harmonic soft spheres~\cite{OSLN03} we choose
\beq\label{eq:Hdef}
H[\vec X] =\sum_{\mu=1}^M v(h_\mu(\vec X)) = \frac12 \sum_{\mu=1}^M h_\mu(\vec X)^2 \th(- h_\mu(\vec X)) \ ,
\qquad\qquad
v(h) = \frac12 h^2 \th(-h) \ .
\eeq
Other choices of $v(h)$ can be considered, provided $v(h>0)=0$ and $v(h<0)>0$. 
The analysis of
the space of solutions can be performed,
following Derrida and Gardner~\cite{DG88}, from the study of the partition function
\beq \label{eq:ZdefG}
Z = \int \DD\vec X e^{-\b H[\vec X]}  = \int \DD\vec X e^{-\frac{\b}2 \sum_{\mu=1}^M h_\mu(\vec X)^2 \th(- h_\mu(\vec X))} \ , 
    \eeq 
    where the measure $\DD\vec X$ may
include some additional normalization constraint.

One is typically interested in the limit $N\to\io$, with the number of constraints $M$ scaled appropriately to have
a non-trivial limit. In this limit, a sharp SAT-UNSAT phase transition emerges~\cite{MZKST99,MPZ02,KMRSZ07}, and can be characterized by looking at
the zero temperature limit of the
partition function.
 In the SAT phase, the ground state energy is equal
to zero, and the partition function reduces to the volume of the
space of satisfying assignments (whose logarithm is the entropy of solutions). 
The corresponding homogeneous
measure on the space of the solutions gives identical weights to all configurations that satisfy the constraints.  
In the UNSAT phase, the ground state
energy is non-zero and the partition function is
dominated by the ground state configurations.

One is usually interested in the free energy per particle,
\beq
\mathrm f = - \frac{T}{N} \overline{ \log Z } \:,
\eeq
where the overline indicates an average over the quenched disorder, if it is present in the constraints.
This quantity has a finite limit for $N\to\io$ and allows one to extract easily all the thermodynamic information;
moreover, in presence of quenched disorder, the free energy per particle is usually
 self-averaging~\cite{MPV87}, i.e. its fluctuations due to the disorder vanish for $N\to\io$.
 In general, ${\rm f} = e - T s$, where $e$ is the thermodynamic energy and $s$ the thermodynamic entropy.
 In particular, in the SAT phase, when $T\to 0$, $e\to 0$ and $s$ has a finite limit; as a result $-\b {\rm f} \to s$.
 On the contrary, in the UNSAT phase, when $T\to 0$, $e$ has a finite limit, and $T s \to 0$, and as a result
 ${\rm f}=e$.
 
 Note that besides the thermodynamic (or ``equilibrium'') SAT-UNSAT transition, defined as above from the partition function in Eq.~\eqref{eq:ZdefG},
one can study many other similar transitions that happen out of equilibrium. Indeed, in the SAT phase at high enough density of constraints, 
the equilibrium state of the system is a collection of distinct thermodynamic states~\cite{KMRSZ07}. One can restrict the study to one of these states (also called
the ``state following'' formalism~\cite{FP95,KZ10,RUYZ15,RU16}) 
and study the SAT-UNSAT transition of the Boltzmann-Gibbs measure restricted to that particular state. It has been shown
for sphere systems that this does not change the critical properties of the transition~\cite{RU16}, hence in the following we restrict our study to the equilibrium
setting.

\subsection{Distribution of gaps}

Besides the free energy and its derived quantities,
another interesting observable, in particular
in the context of jamming, is the probability distribution of ``gaps''. In fact, it has been shown that this
quantity encodes important information about the marginal stability of jammed configurations that we are 
going to discuss below~\cite{Wy12}.

A gap variable is just a constraint function $h_\mu(\vec X)$ -- the name ``gap'' originates from the fact that
$h_\mu=0$ corresponds to a constraint being on the verge of becoming unsatisfied (or a ``contact''), 
and the value of $h_\mu$ is thus the distance (``gap'') to this configuration.
The gap probability distribution $\r(h)$ is defined as
\beq
\r(h)= \overline{\langle \hat \r(h)\rangle} \ , \ \ \ \ \ \ \ \ \ \hat\r(h) = \frac1M \sum_{\m=1}^M \d(h-h_\m(\vec X)) \ ,
\eeq
where the brackets denote a thermodynamical average, while the overline denotes an average over quenched disorder,
if present.
From $\r(h)$ one can derive 
\beq\label{eq:zdef}
z\equiv \int_{-\infty}^0 \de h\, \r(h)\:,
\eeq
which is the average fraction of unsatisfied constraints, or fraction of contacts.
To compute the distribution of gaps, we note that the partition function can be written as
\beq
Z = \int \DD\vec X e^{-\b H[\vec X]} = \int \DD\vec X e^{M \int \de h \hat\r(h) [-\b v(h)]} \:.
\eeq
Hence, because $-\b {\rm f} = \overline{\log Z }/N$,
we get
\beq\label{eq:rhohformula}
\frac{\de {\rm f}}{\de v(h)} = 
\frac1N \frac{\de\, \overline{ \log Z}}{\de [-\b v(h)]} = \a  \overline{ \la \hat\r(h) \ra } = \a \, \r(h)\:.
\eeq
In the SAT phase, there are no unsatisfied constraints, and $\r(h)=0$ for $h<0$, so that $z=0$, while in the UNSAT phase
$\r(h)$ is non-zero for $h<0$ and $z>0$. A particularly important quantity is the limit value of $z$ when one approaches
the SAT-UNSAT transition coming from the UNSAT phase where $z>0$.
This limit is usually strictly positive, hence $z$ jumps discontinuously
at the transition. If this limiting value is such that the number of unsatisfied constraints is exactly equal to the number of degrees of freedom,
i.e. $M z=N$, then the system is said to be {\it isostatic}~\cite{LNSW10}.
We can therefore define an {\it isostaticity index} $c = (M/N) z$, which is equal to 1 if the system is isostatic.
More generally,
the system is said to be hypostatic, isostatic or hyperstatic 
whenever the total number of violated constraints is less ($c<1$), equal ($c=1$) or 
higher ($c>1$) than the total number of degrees of freedom.

We will be particularly interested in observables of the form
$\OO[\vec X] = M^{-1}\sum_{\mu=1}^M \OO(h_\mu) \th(-h_\mu)$, that 
are functions of the negative gaps. We define the thermodynamic and disorder average
\beq\label{eq:OO1}
[\OO(h)] = 
\overline{\la \OO \ra} \equiv
\frac 1M \sum_{\mu=1}^{M} \overline{\langle\OO(h_\mu)\th(-h_\mu)\rangle} =
 \int \de h \r(h) \OO(h) \th(-h)  \ .
\eeq
Special cases of this class of observables are the thermodynamic energy $e = \overline{\la H[\vec X]\ra}/N$, the ``pressure'' $p$, and the fraction of contacts:
\beq\label{eq:zpe}
z = [1] \ ,
\qquad
p = - [h] \ ,
\qquad
e = \a[v(h)] =\a [h^2]/2 \ .
\eeq

\subsection{Some consequences of isostaticity: force distribution and soft modes}

We now formulate in our abstract continuous CSP setting some well-known
consequences of isostaticity~\cite{LDW13,CCPZ15}, i.e. the condition that the number of unsatisfied constraints is equal to the number of degrees of freedom, and $c = 1$.
We omit for notational simplicity the $\vec X$-dependence of the gaps $h_\mu$ and we define, with respect to the Hamiltonian given in Eq.~\eqref{eq:Hdef}:
\beq\label{eq:fmudef}
\begin{split}
F_i &= - \frac{\de H}{\de X_i} = \sum_{\m=1}^M [ -h_\mu \th(-h_\mu) ] \frac{\de h_\mu}{\de X_i}
= \sum_{\mu \in \CC} \SS_{\m i} f_\mu
 \ , \\
 f_\mu &= - h_\mu \th(-h_\mu) \ , \\
 \SS_{\mu i} &= \frac{\de h_\mu}{\de X_i} \ ,
\end{split}\eeq
where the set of ``contacts''
$\CC = \{ \mu : h_\mu < 0\}$ is the set of unsatisfied constraints, of size $|\CC| = M z = N c$,
the matrix $\SS$ has dimension $M z \times N$, and the ``contact force'' vector $\vec f$ has dimension $M z$ because we consider only the non-zero components $f_\m$.
We therefore have $\vec F = \SS^T \vec f$. 
At zero temperature, we are especially interested in minima of the Hamiltonian, which correspond to vanishing ``total forces'' $\vec F = \SS^T \vec f = \vec 0$.
Furthermore, one is often interested in the small harmonic vibrations around a minimum, which are described by the Hessian matrix
\beq\label{eq:Hessian}
\HH_{ij} = \frac{\de ^2 H}{\de X_i \de X_j} = \sum_{\m\in \CC} \frac{\de h_\mu}{\de X_i} \frac{\de h_\mu}{\de X_j} +\sum_{\mu\in \CC} h_\m  \frac{\de ^2 h_\m}{\de X_i \de X_j} \ .
\eeq

It is particularly interesting to consider the above structure in the jamming limit, i.e. at the SAT-UNSAT transition. Note that ``contacts'', 
i.e. constraints for which $h_\mu<0$ in the UNSAT phase close to the transition, must then become marginally satisfied, i.e.
$h_\mu=0$, right at the transition. 
Moreover, the average contact force is proportional to the pressure, $|\CC|^{-1} \sum_{\m\in \CC} f_\m = -[h]/[1] = p/[1]$, which indeed vanishes at jamming.
For contacts, one can thus define scaled forces $f^s_\mu = [1] f_\mu/p$, that remain finite at the jamming transition. These scaled forces still satisfy the condition 
$\SS^T \vec f^s =0$, where the matrix $\SS$ is calculated at the jamming point. 
Although by definition both $\SS$ and $\vec f$ are fully determined by the configuration $\vec X$, one can ask in general how many solutions $\vec f$ to
$\SS^T \vec f = \vec 0$ can be found, for fixed $\SS$. Because this is a linear equation, for hypostatic systems ($c<1$) there are in general no solutions,
while for hyperstatic systems ($c>1$) there are in general $N(c-1)$ solutions, and for isostatic systems ($c=1$) there is a single solution. Therefore, for an isostatic system, 
the vector of scaled forces must necessarily coincide with the unique solution of $\SS^T \vec f^s = \vec 0$. This is particularly useful for numerical calculations, because the
force vector vanishes at jamming and it is therefore difficult to evaluate the scaled forces with high precision, while the matrix $\SS$ remains finite and determining its unique
zero mode requires much lower precision~\cite{CCPZ15}. Moreover, while in the SAT phase in principle the contact forces vanish, one can still define effective contact forces
by following the procedure outlined in~\cite{BW06}. Then, one finds that the scaled contact forces also converge to the zero mode of $\SS$ when approaching jamming from
the SAT phase.
Finally, one can consider the Hessian matrix at jamming. The second term in Eq.~\eqref{eq:Hessian} vanishes because $h_\m=0$ for contacts, and therefore
$\HH = \SS^T \SS$. From this one can deduce that in general $\HH$ has $N(1-c)$ zero modes for hypostatic systems, while it has no zero modes for hyperstatic systems;
therefore at jamming, which separates the two situations, one finds a large number of very small eigenvalues of the Hessian (``soft modes'')~\cite{LDW13}.

We have seen in this section that, for general continuous CSPs, isostaticity at jamming implies that the contact forces are fully determined by the matrix $\SS$, that also determines
the soft modes of the Hessian. This mathematical structure has many other interesting consequences.
We will not provide further details on these aspects. The interested reader can consult Ref.~\cite[Supplementary Information]{CCPZ15} for a detailed review of the properties of $\SS$ 
in the context of spheres, 
Ref.~\cite{FPUZ15} for a study of the Hessian in the perceptron, and Ref.~\cite{Wy12,LDW13} for a discussion of soft modes in sphere packings.

\subsection{Sphere packing as a constraint satisfaction problem}

In order to motivate the interest for the observables discussed above (and their names), we discuss here more explicitly the
sphere packing problem and its formulation as a continuous CSP.
The $d$-dimensional sphere packing problem is indeed a special case of the general setting discussed above.
One considers $n$ points in a $d$ dimensions volume, $\bx_i \in V \subset \RRR^d$, $i=1,\cdots,n$,
hence the total number of degrees of freedom is $N=d n$.
There are $M = n(n-1)/2$ constraints corresponding to all possible distinct particle pairs $\mu = \la i,j \ra$ (e.g. with $i<j$),
of the form
\beq
h_\mu(\vec X) = | \bx_i - \bx_j | - \s_{ij} \ ,
\qquad
\s_{ij} = \frac{\s_i + \s_j}2 \ .
\eeq
In this case, $\s_i$ can be interpreted as the diameter of particle $i$, and then $h_{\la i,j\ra}$ is precisely the physical gap
between particles $i,j$. In particular, if $h_{\la i,j\ra}<0$, then the two particles $i,j$ overlap and thus feel a repulsive interaction.
This justifies the name ``gap'' given to $h_\mu$, and the name ``fraction of contacts'' given to $z$.
Note that the fraction of contacts defined in Eq.~\eqref{eq:zdef} is normalized to the total number of constraints; in particle systems,
it is customary to consider instead the average number of contacts per particle,
\beq
z_p = \frac{2 M z}{n} = \frac{2 N c}{n} = 2 d \, c \ .
\eeq
Hence, an isostaticity index $c=1$ corresponds to $z_p=2d$ contacts per particle.
The Hamiltonian $H[\vec X]$ in Eq.~\eqref{eq:Hdef}, with the choice $v(h) = h^2 \th(-h)/2$, is precisely the Hamiltonian 
of a system of soft harmonic repulsive spheres~\cite{OSLN03,LNSW10}; the partition function in Eq.~\eqref{eq:ZdefG} is the thermodynamical
canonical partition function of the model, and $\rm f$ the associated free energy per particle.
The distribution of gaps is simply related to the ``radial distribution function'' of the particle system, and the average gap is related to the pressure.
Finally, $f_\mu$ defined in Eq.~\eqref{eq:fmudef} is precisely the modulus of the contact force associated to the particle pair $\la ij \ra$, and the matrix
$\SS$ encodes the network of contacts in the particle packing~\cite{CCPZ15}.
In this case, one is then interested in the limit $n,|V|\to\io$ with fixed density $n/|V|$.

Note that in this case, in the SAT phase at $T\to 0$, the Boltzmann-Gibbs measure becomes a uniform measure over all the configurations
satisfying the non-overlapping constraint, which coincides with the equilibrium Boltzmann-Gibbs distribution
of a system of hard spheres. In the UNSAT phase
(which cannot exist for hard spheres) there are overlaps and, at zero temperature, one has instead a mechanically stable assembly
of soft repulsive spheres.

In the sphere packing problem, at least when $\s_{ij} = \s$ (all particles are identical), there is an additional complication because, even within
the SAT phase, the system can form a {\it crystal}, a phase in which the particles are arranged over a regular lattice in $\RRR^d$. Crystals
are usually denser than disordered arrangements, and therefore the SAT-UNSAT transition usually happens within the crystal phase,
although the situation is somewhat uncertain in large dimensions~\cite{TS10}. In this case, the SAT-UNSAT transition (also called ``close packing'' point)
has very different properties with respect to the same transition in disordered assemblies. Yet, one can restrict to the study of disordered configurations
and in that case the SAT-UNSAT transition (also called ``jamming transition'' or ``random close packing'' in this case) 
has robustly universal critical properties~\cite{PZ10,LNSW10}.

\section{Definition of the model and basic equations}
\label{sec:perc_def}

Following~\cite{FP15}, in this paper we want to study the properties of the simplest continuous CSP that exhibits a phenomenology similar
to that of spheres (once one restricts to their disordered phase). While spheres do not have quenched disorder (the constraints are deterministic functions),
the perceptron model has quenched disorder. The presence of quenched disorder eliminates any possible crystal phase and one can then study the
equilibrium properties of the model in the disordered phase.
Before getting into the study of the phase diagram, and its critical properties at the SAT-UNSAT transition, in this section we give
the basic definition of the model and the main equations needed for its study.

\subsection{Definition of the model and free energy}
\label{basics}

The perceptron model, on which we concentrate in the rest of this paper, is defined by the
linear functions 
\begin{eqnarray}
  \label{eq:13}
h_\mu(\vec X) = \frac1{\sqrt{N}} \vec X  \cdot \vec \xi^\mu - \s 
\end{eqnarray}
with the normalization $\vec X\cdot \vec X = N$. The vectors
$\vec\x^\m$, that following the neural network literature~\cite{DG88} we call
``patterns'', have components $\xi_i^\mu$ which are independent
Gaussian variables of zero average and unit variance. 
The partition function reads
\beq\label{eq:Zdef} 
Z = \int \DD\vec X e^{-\b H[\vec X]} = \int \DD\vec X \prod_{\m=1}^M \left[ \int
  \de r_\m e^{-\b v(r_\m-\s)} \d\left(r_\m - N^{-1/2} \vec X \cdot \vec
    \xi^\mu \right) \right] \ , 
    \eeq 
where the measure $\DD\vec X$ contains the constraint $\vec X\cdot \vec X = N$,
i.e. it is the uniform measure on the $N$-dimensional sphere of radius $\sqrt{N}$.
 For positive
$\sigma$ the model can be interpreted as a classifier of the random
patterns $\vec\xi^\mu$, and defines a convex optimization problem. For
negative $\sigma$ the model cannot be interpreted as a classifier. It
still is a legitimate CSP, but it is non-convex. It has been
observed in~\cite{FP15} that this is the interesting regime to
describe jamming and glassy phases. In both cases one considers 
the thermodynamic limit $N,M \to\infty$ for values of $\alpha=M/N$ fixed. 

The disorder average of the free energy can be computed via the replica method:
\beq
\mathrm f = - \frac{T}{N} \overline{ \log Z } = - \frac{T}{N}  \lim_{n\to 0}  \partial_n \overline{Z^n} \:.
\eeq
The free energy $\rm f$ is obtained through standard manipulations \cite{MPV87} from
the expression of $Z^n$ for integer $n$. At the end of a computation
sketched in Appendix \ref{sec:App1}, the free energy
can be written as a saddle point over the $n\times n$ ``replica overlap matrix'' 
\beq
Q_{ab}=\frac1N \overline{\langle \vec{X_a}\cdot \vec{X_b}\rangle} \ ,
\eeq
where $\vec X_{a}$ are
replicas of the configurations of the system. Notice that the spherical
contraint on the $\vec X$ implies $Q_{aa}=1$. 
The explicit expression for the
free-energy is
\beq
\mathrm f = -T \lim_{n\to 0} s.p.\partial_n S(Q) \ ,
\eeq
where $s.p.$ denotes the saddle point over the values $Q_{ab}$ and 
\beq\label{eq:SQdef}
\begin{split}
 S(Q) = \frac12 \log\det Q + \a \log \left( \left.
e^{\frac12 \sum_{ab}^n Q_{ab} \frac{\partial^2}{\partial h_a \partial h_b}}
\prod_{a=1}^n e^{-\b v(h_a)}
\right|_{h_a=-\s} \right)
\end{split}\eeq
is the \emph{replicated free energy}.
The saddle point equation for $Q_{ab}$ is thus
\beq
0=\left[Q^{-1}\right]_{ab} + \a \left(\b^2 \langle v'(h_a)v'(h_b)\rangle -\b \d_{ab}\langle v''(h_a)\rangle\right)
\eeq
where we have defined
\beq
\langle O(h_a,h_b,\ldots)\rangle=\frac{\left.e^{\frac12 \sum_{ab}^n Q_{ab} \frac{\partial^2}{\partial h_a \partial h_b}}
O(h_a,h_b,\ldots)\prod_{a=1}^n e^{-\b v(h_a)}\right|_{h_a=-\s}}{\left.
e^{\frac12 \sum_{ab} Q_{ab} \frac{\partial^2}{\partial h_a \partial h_b}}
\prod_{a=1}^n e^{-\b v(h_a)}
\right|_{h_a=-\s}} \ .
\label{SP_general}
\eeq
Finding the solution of such kind of equations is an extremely difficult task.
Furthermore, once the solution is found a sensible analytic continuation down to $n\to 0$
must be taken, which is an additional complication.
The solution to both difficulties is provided by the use of hierarchical matrices. 
Here we do not review in detail this construction,
which can be found in several reviews, e.g.~\cite{MPV87}.

\subsection{Hierarchical ansatz for the saddle point solution}
\begin{figure}[h]
\includegraphics[width=.4\textwidth]{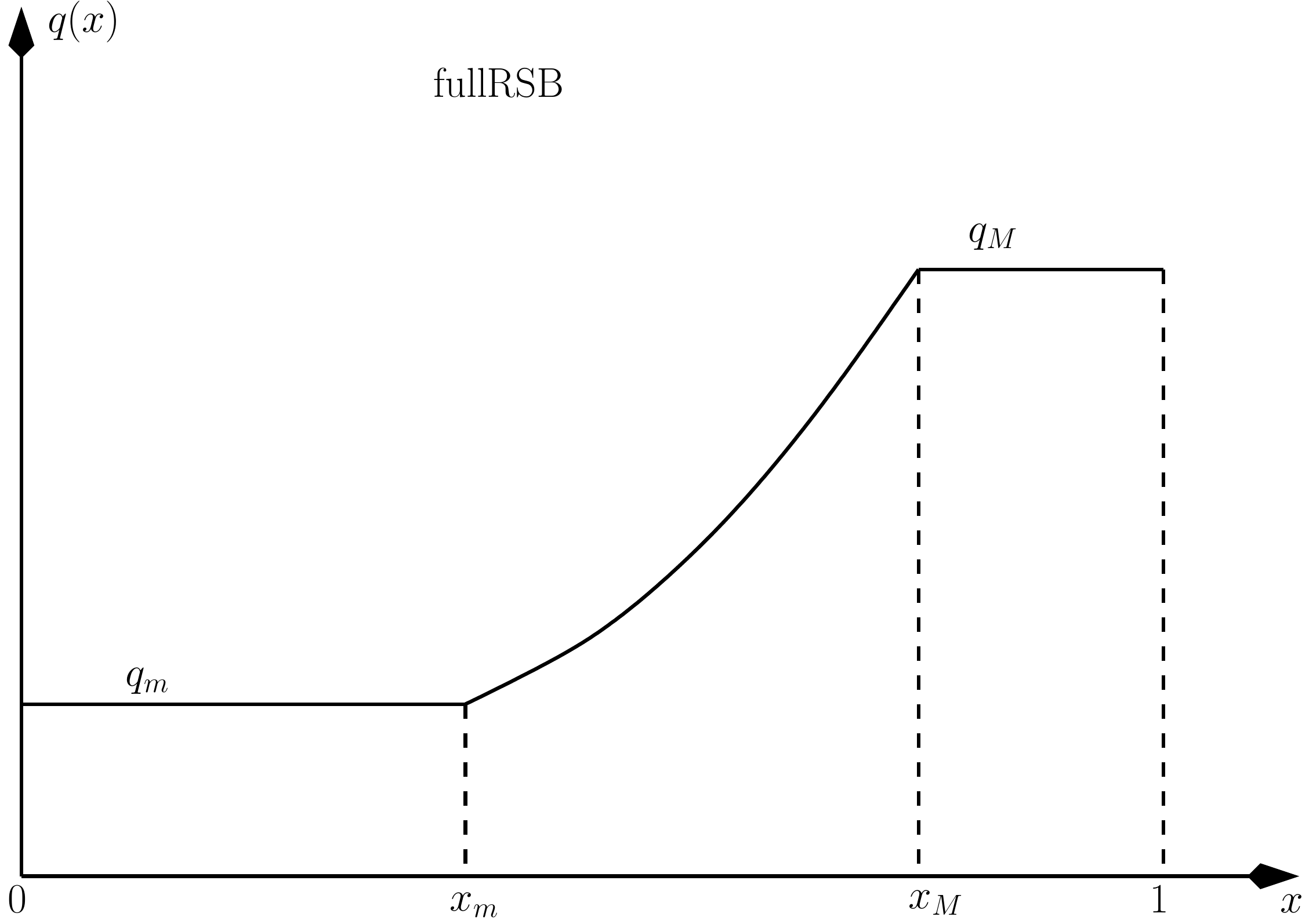}
\caption{The expected form of the function $q(x)$ in the fullRSB phase.
}
\label{fig:hier}
\end{figure}

The general solution of Eq.~(\ref{SP_general}), in the limit $n\to 0$, 
is given in terms of a continuous Parisi hierarchical matrix.
In this case the matrix $Q_{ab}$ is parametrized by a function $q(x)$ in
the interval $x\in [0,1]$ whose general form is plotted in Fig.~\ref{fig:hier}.
For a hierarchical matrix, the first term of $S(Q)$ reads~\cite{MP91}
\begin{eqnarray}
  \label{eq:2}
  \lim_{n\to 0}\frac 1 n \log\det
  Q=\log(1-q_M)+\frac{q_m}{\l(0)}+\int_0^1 \de x\;
  \frac{\dot q(x)}{\lambda(x)} \ ,
\end{eqnarray}
where the dot denotes differentiation with respect to $x$, and
\beq
\l(x)=1-xq(x)-\int_{x}^1\de y\,  q(y)\:.
\eeq
Note that $\dot\l(x)= -x\dot q(x)$, and therefore when $q(x)$ is constant also $\l(x)$ is constant;
furthermore, $\l(0) = \l(x_m) = 1 - \int_0^1 \de x q(x)$, while $\l(1) = \l(x_M) = 1 - q_M$.
The second term can be written as \cite{Du81}
\begin{eqnarray}
  \label{eq:3}
 \lim_{n\to 0} \frac1n \log \left.
e^{\frac12 \sum_{ab} Q_{ab} \frac{\partial^2}{\partial h_a \partial h_b}}
\prod_a e^{-\b v(h_a)}
\right|_{h_a=-\s} =\gamma_{q_m}\star f(0,h) |_{h=-\sigma}
\end{eqnarray}
where we have defined the convolution
\beq
\gamma_q\star g(h)=\int_{-\infty}^\infty \frac{\de y}{\sqrt{2\pi q}}
e^{-\frac{y^2}{2q}}g(h-y).  
\eeq
The function $f(x,h)$ verifies the Parisi equation (the dot represents a $x$-differentiation, while the prime a $h$-differentiation):
\beq
\dot f(x,h)= -\frac 12\dot q(x) \left[ f''(x, h) + x f'(x, h)^2    \right] \ , \hskip10pt x_m < x < x_M\:,
\label{eqf_x}
\eeq
while for $x \notin [x_m,x_M]$ one has $\dot q(x)=0$ and $\dot f(x,h)=0$, hence $f(x,h)$ is independent of $x$.
The boundary condition for $f(x,h)$ at $x=1$ (or, equivalently, at $x=x_M$) is:
\begin{eqnarray}
  \label{eqf_x_i}
f(1,h) = \log \g_{1 - q_M} \star e^{-\b v(h)} \ .
\end{eqnarray}

It is useful to introduce the inverse function of $q(x)$, called
$x(q)$, which is defined for $q\in[q_m,q_M]$ where $q_m$ and $q_M$ are defined
in Fig.~\ref{fig:hier}.  Defining $f(q,h)\equiv f(x(q),h)$,
Eqs.~(\ref{eqf_x},\ref{eqf_x_i}) become
\beq\label{eqf}
\begin{cases}
f(q_M,h) = \log \g_{1 - q_M} \star e^{-\b v(h)} \ , \\
\dot f(q,h)= -\frac 12 \left[ f''(q, h) + x(q) f'(q, h)^2    \right] \ , \hskip10pt q_m < q < q_M \:. 
\end{cases}
\eeq
where now the dot represents $q$-differentiation. 
Thus, for a given function $x(q)$ and $q_m$ and $q_M$ we can solve Eq.~(\ref{eqf}) to compute $f(q,h)$. 
Then the replicated free energy computed on this particular function is
\beq
  -\b {\rm f}[x(q)]\equiv \lim_{n\to 0}\partial_n S(Q)=\frac 1 2 \left[
\log(1-q_M)+\frac{q_m}{\l(q_m)}+\int_{q_m}^{q_M}
  \frac{\de q}{\lambda(q)}
\right]
+\alpha \gamma_{q_m}\star f(q_m,h) |_{h=-\sigma}
\label{fe}
\eeq
where we have introduced $\l(q) = \l(x(q))$, given by
\beq\label{eq:ql}
\l(q) =  1 - q_M + \int_q^{q_M} \de p \, x(p) \ .
\eeq
The next step is to find the variational equations for the function $x(q)$, or equivalently $q(x)$.

\subsection{Variational equations}\label{sec:fullRSBeq}

In order to obtain the equations that determine the function $x(q)$ we need to impose that the function $f(q,h)$ that appears in Eq.~(\ref{fe})
satisfies Eq.~(\ref{eqf}).
A simple way to impose Eq.~(\ref{eqf}) is to add a Lagrange multiplier $P(q,h)$ to $s[x(q)]$. The new variational free energy thus becomes
\begin{eqnarray}
  \label{eq_s}
  -\b {\rm f}[x(q)]&=&\frac 1 2 \left[
\log(1-q_M)+\frac{q_m}{\l(q_m)}+\int_{q_m}^{q_M}
  \frac{\de q}{\lambda(q)}
\right]
+\alpha \gamma_{q_m}\star f(q_m,h) |_{h=-\sigma} \nonumber \\
&&
-\alpha \int \de h\; P(q_M,h)\;[f(q_M,h) - \log \g_{1 - q_M} \star e^{-\b
  v(h)}]\\
&&
+\alpha  \int \de h\;\int_{q_m}^{q_M} \de q\;  P(q,h)\left\{ \dot f(q,h)+\frac 12 \left[ f''(q, h) + x(q) f'(q, h)^2    \right]  \right\} \nonumber \ .
\end{eqnarray}

Taking the variational equation with respect to $P(q_M, h)$ and $P(q,h)$ gives back Eq.~(\ref{eqf}).
Now we can take the variational equations with respect to
$f(q,h)$, $f(q_m,h)$, and $x(q)$~\cite{MPV87}.
The resulting equations are:
\beq\label{eq:qP}
\begin{cases}
P(q_m,h)  = \g_{q_m}(h+\s) \\
\dot P(q,h) = \frac12 \left[P''(q,h)-2x(q) (P(q,h)f'(q,h))'\right] \ , \hskip10pt q_m < q < q_M 
\ ,
\end{cases}
\eeq
and
\beq\label{eq:qx1}
\frac{q_m}{\l(q_m)^2} + \int_{q_m}^q \de p \frac1{\l(p)^2} = \a \int \de h P(q,h) f'(q,h)^2 \ .
\eeq
Note that $P(q,h)$ is normalised to 1 for all $q$. Whenever
$x(q)$ has a continuous part (i.e. $\dot x(q)\neq 0$), one can differentiate
Eq.~(\ref{eq:qx1}) w.r.t. $q$; the first and second 
derivatives lead, respectively, to explicit expressions of $\lambda(q)$ and
$x(q)$ as functions of $f(q,h)$ and $P(q,h)$:
\beq\label{eq:qx2}
\frac1{\l(q)^2} = \a \int \de h P(q,h) f''(q,h)^2 \ ,
\eeq
and
\beq\label{breaking_point}
x(q) = \frac{\l(q)}2 
\frac
{ \int \de h P(q,h) f'''(q,h)^2  } 
{ \int \de h P(q,h) [ f''(q,h)^2 + \l(q) f''(q,h)^3 ] } \:.
\eeq
We stress once again that Eqs.~\eqref{eq:qx2} and \eqref{breaking_point}
only hold whenever $\dot x(q) \neq 0$.

\subsection{Iterative solution of the saddle point equations}
\label{sec:num_sol}

The thermodynamic value of the free-energy at given values of the
parameters $(\sigma,\alpha)$ can be obtained from the numerical
solution of the variational equations. This can be obtained by iteration
according to the following procedure:
{\tt
\begin{enumerate}
\item Start with a guess for $x(q)$, $q_m \leq q \leq q_M$;
\item Use Eqs.~\eqref{eqf} and \eqref{eq:qP} to obtain an estimate of $f(q,h)$ and $P(q,h)$;
\item Get $q_m$ from the ratio of Eqs.~\eqref{eq:qx1} and \eqref{eq:qx2}, computed in $q=q_m$;
\item Obtain a new guess for $q_M$ from  Eqs.~\eqref{eq:qx2} and \eqref{eq:ql} computed in $q=q_M$;
 \item Use Eq.~\eqref{eq:ql} to obtain $\l(q)$
 and Eq.~\eqref{breaking_point} to obtain $x(q)$;
\item Iterate from 2 until convergence.
\end{enumerate}} 
The numerical solution of the partial differential equations
(\ref{eqf}) and (\ref{eq:qP}) can be obtained by discretizing the
profile $x(q)$ through a stepwise function with $K$ steps (this is
known as the $K$RSB approximation) and then increasing $K$ until convergence. 
The details of this procedure are
described in Appendix \ref{sec:KRSB}.

\subsection{Distribution of gaps and contacts}

From the general relation in Eq.~\eqref{eq:rhohformula}, and
using Eq.~\eqref{eq_s} we get
\beq\label{eq:rhoh}
\r(h) = \frac1\a \frac{\de {\rm f}}{\de v(h)} =e^{-\b v(h)} \int \de z P(q_M,z) e^{-f(q_M,z)} \g_{1-q_M}(z-h) \ .
\eeq
Note that $\r(h)$ is correctly normalized so that $\int \de h \r(h)=1$. 
Note also that for an observable that is a function of the gaps, 
we have from Eq.~\eqref{eq:OO1}:
\beq\label{eq:OO}
[ \OO ] = \int \de h \r(h) \OO(h)\th(-h) = \int \de h P(q_M,h) \frac{ \g_{1-q_M}\star [ e^{-\b v(h)} \OO(h)\th(-h) ] }{\g_{1-q_M}\star e^{-\b v(h)}  } \ .
\eeq
In particular the fraction of contacts is given by
\beq\label{eq:zperc}
z=\int \de h\, \r(h) \theta(-h) = \int \de h\, P(q_M,h) \frac{ \g_{1-q_M}\star [ e^{-\b v(h)} \theta(-h) ] }{\g_{1-q_M}\star e^{-\b v(h)} }\:.
\eeq

\section{The zero temperature phase diagram}
\label{sec:PD}

The formulae derived in Sec.~\ref{sec:perc_def} hold for any value of the control parameters:
the density of constraints $\a$, the parameter $\s$ that enters into the constraints,
and the inverse temperature $\b$. As a function of these control parameters, the order parameter function $q(x)$ has different forms,
giving rise to several distinct phases and phase transitions. 
To simplify the study of this phase diagram, here 
we specialise to the zero temperature limit where a sharp SAT-UNSAT (or, equivalently, jamming) phase transition is found. Note that at finite temperature there is always a finite
probability of violating some constraint, and the transition is smoothed out.
The zero temperature phase diagram of the perceptron
has been discussed for $\s\geq 0$ in~\cite{DG88}, and for $\s<0$ (but $|\s|$ not too large) in~\cite{FPUZ15}. 
Here, we discuss the complete phase diagram for all values of $\s$ and $\a$, with the result given in Fig.~\ref{PD},
and characterize all the phases and phase transitions that appear.

\subsection{The replica symmetric ansatz}

The solution of replica equations like the ones derived in Sec.~\ref{sec:perc_def} is usually obtained through a series of steps. 
One starts by the simplest possible solution, 
called the ``replica symmetric'' (RS) solution.
It corresponds to a constant $q(x)=q_M$, in which case the matrix $Q_{ab} = q_M$ for all $a\neq b$.
Because $\dot q(x)=0$, one also has $f(q,h)=f(q_M,h)=\log \g_{1 - q_M} \star e^{-\b v(h)}$ and $P(q,h)=P(q_M,h)=\g_{q_M}(h+\s) $.
Thus, at the RS level we have
\beq
-\b {\rm f}_{RS}(q_M) =\frac 1 2 \left[
\log(1-q_M)+\frac{q_M}{1- q_M}\right]
+\alpha \gamma_{q_M}\star f(q_M,h) |_{h=-\sigma}\:.
\label{s_RS}
\eeq
When we take the zero temperature limit of this expression, we need to specify whether we are in a SAT or UNSAT phase.
We should note that the RS ansatz amounts to assume that the space of solutions form a unique connected component,
or equivalently that the free energy has a single minimum~\cite{MPV87,KMRSZ07}.
In the SAT phase the value of $q_M$ remains finite for $T\to 0$: because $q_M$ measures the similarity between two solutions,
when the volume of the space of 
solutions is finite, two typical solutions differ and thus $q_M<1$. 
Instead, in the UNSAT phase, under the assumption that there is a unique minimum, all the replicas converge towards the same
state when $T\to 0$ and therefore one has $q\to 1$ in that limit.

\subsubsection{The SAT phase}

Taking the zero temperature limit with constant $q_M < 1$, we get
\beq\label{eq:fSAT}
\begin{split}
f(q_M,h)&=\lim_{\b\to \infty}\log \g_{1-q_M}\star e^{-\b v(h)} = \log \g_{1-q_M}\star \theta(h) \\
&=\log \int_{-\infty}^h \frac{\de z}{\sqrt{2\pi(1-q_M)}}e^{-\frac{z^2}{2(1-q_M)}}\equiv \log \Theta\left(\frac{h}{\sqrt{2(1-q_M)}}\right)
\equiv f_{\mathrm{SAT}}(q_M, h)
 \ ,
\end{split}
\eeq
where
\beq
\Theta(x)=\frac 12\left(1+\erf(x)\right)\:.
\eeq
Note that in this case the function $-\b {\rm f}_{RS}(q_M)$ in Eq.~\eqref{s_RS} has a finite limit for $T\to 0$, which gives the entropy of the system, 
$s(q_M) = \lim_{T\to 0} [-\b {\rm f}_{RS}(q_M)]$,
i.e. the logarithm of the volume of the space of solutions.

The saddle point equation for $q_M$ can be obtained either by taking the derivative of  Eq.~(\ref{s_RS}) with respect to $q_M$ or by considering explicitly the RS ansatz in Eq.~(\ref{eq:qx1}). In the last case we obtain
\beq
\frac{q_M}{(1-q_M)^2}=\a \int_{-\infty}^\infty \frac{\de h}{\sqrt{2\pi q_M}}e^{-\frac{(\s+ h)^2}{2 q_M}} \left[\frac{\de }{\de h}\log \Th\left(\frac{h}{\sqrt{2(1-q_M)}}\right)\right]^2\:.
\label{RS_sp}
\eeq
One can solve Eq.~\eqref{RS_sp} numerically, and in
the SAT phase it is found, as expected, that $q_M<1$. 
Within this solution, the SAT-UNSAT transition point is reached when $q_M\to 1$. 
Taking this limit in Eq.~(\ref{RS_sp}), using the asymptotic properties of the error function,
we get the equation for the critical satisfiability threshold $\a_J(\sigma)$ within the replica symmetric ansatz:
\beq\label{eq:alphaJRS}
\a_{\mathrm{J}}(\s)=\left[\int_{-\infty}^\s \frac{\de h}{\sqrt{2\pi}}e^{-\frac{h^2}2}(h-\s)^2\right]^{-1} \ .
\eeq
For $\s>0$, our optimization problem is convex, the RS solution is always stable (see Sec.~\ref{sec:RSstab}), and Eq.~\eqref{eq:alphaJRS}
gives the correct result for the SAT-UNSAT transition line~\cite{DG88}, as shown in Fig.~\ref{PD}.

\subsubsection{The UNSAT phase}

Within the RS ansatz, in the UNSAT phase, there is a unique energy minimum and the free energy converges to its energy. 
Correspondingly, $q_M \to 1$. At very low temperature, the system performs harmonic vibrations around that minimum, and in that case
one can show that $q_M=1-\chi T + O(T^2)$. We thus take the $T\to 0$ limit with $1-q_M = \chi T \to 0$ at the same time.
Plugging this scaling in $f(q_M,h) = \log \g_{1 - q_M} \star e^{-\b v(h)}$, with $v(h) = h^2 \th(-h)/2$, we get at leading order in $\b$ (see Appendix \ref{App_Asy})
\beq\label{eq:fasyunsat}
f(1-\chi /\b,h)\simeq -\frac{\b h^2}{2(1+\chi)}\theta(-h) \ ,
\eeq
Therefore, the ground state energy is
\beq
e_{RS} = \lim_{T\to 0}{\rm f}_{RS}(q_M=1-\chi T) = -  
\frac{1}{2\chi}
+ \frac{\alpha}{2(1+\chi)} \int_{-\infty}^\s \frac{\de h}{\sqrt{2\pi}}e^{-\frac{h^2}2} (h-\s)^2 \:.
\label{e_RS}
\eeq
and the saddle point equation for $\chi$ is given by
\beq
\frac{1}{\chi^2}=\frac{\a}{(1+\chi)^2}\int_{-\infty}^\s \frac{\de h}{\sqrt{2\pi}}e^{-\frac{h^2}2}(h-\s)^2 
\qquad
\Leftrightarrow
\qquad
\left(1 + \frac1\c \right)^2 = \frac{\a}{\a_J(\s)} \ ,
\label{RS_unsat}
\eeq
where $\a_J(\s)$ is given in Eq.~\eqref{eq:alphaJRS}. Eq.~\eqref{RS_unsat} has a solution for $\c$ only for $\a > \a_J(\s)$, which
indeed defines the UNSAT phase.
In the SAT phase, $q_M$ remains less than one for $T\to 0$, while in the UNSAT phase we have $q_M = 1 -\chi T$.
To match the two scalings, when $\a \to \a_J(\s)$ from the UNSAT phase, we should have that $\chi\to \infty$,
which indeed follows from Eq.~\eqref{RS_unsat}.
Furthermore, inserting the saddle point equation~\eqref{RS_unsat} in the expression of the energy Eq.~\eqref{e_RS}, we obtain
\beq
e_{RS} = \frac1{2\chi^2}  = \frac12 \left( \sqrt{\frac{\a}{\a_J(\s)}} - 1 \right)^2 
\ .
\eeq
For fixed $\s$ and $\a=\a_{\mathrm J}(\s)+\delta \a$,
Eq.~(\ref{RS_unsat}) shows that $\chi^{-1}\sim \delta \alpha$ when $\d\a\to 0$,
and therefore, close to the jamming line, the energy goes as $e\sim \delta \a^2$. A similar result is obtained at fixed $\a$ as a function of $\d\s$: the energy always
scales as the square of the distance from the jamming line.

\begin{figure}[t]
\centering
\includegraphics[width=.7\textwidth]{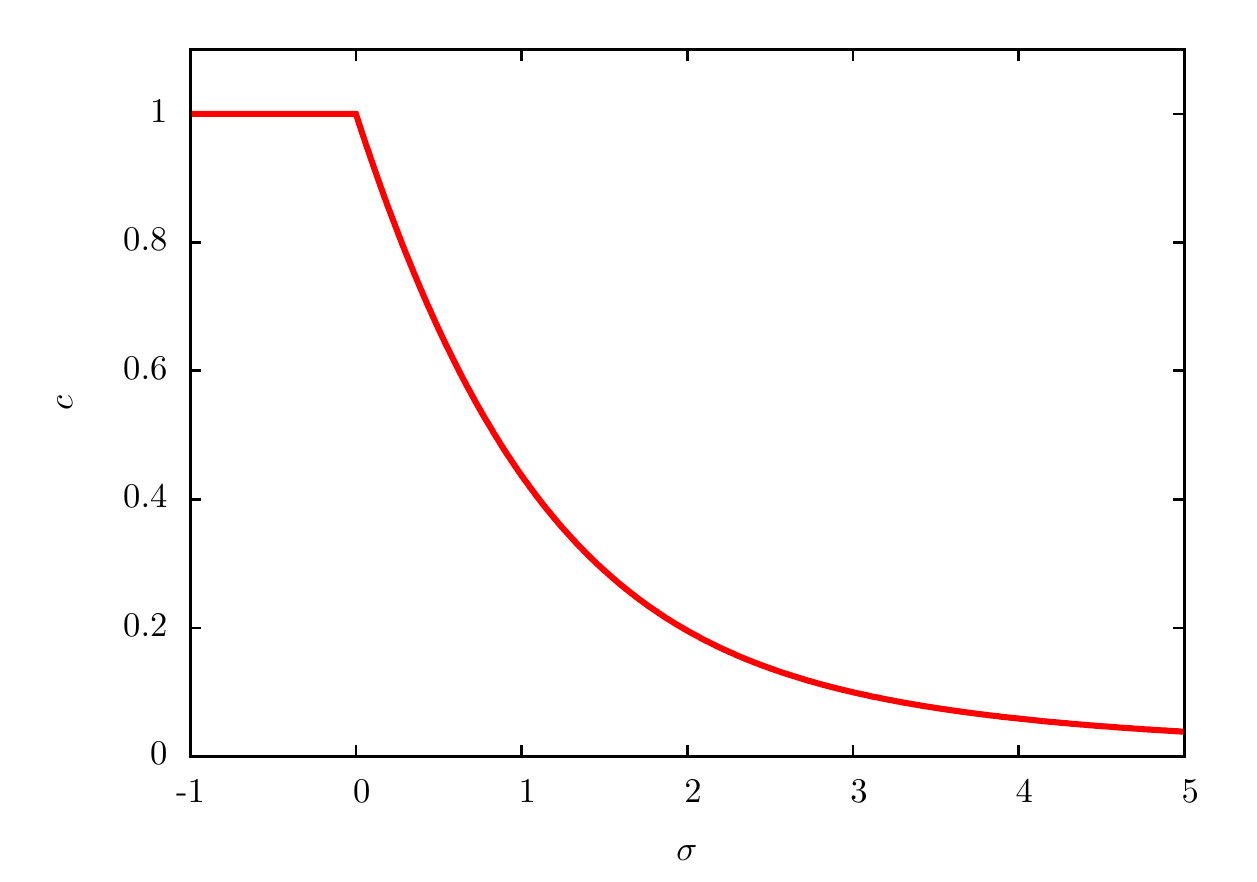}
\caption{Isostaticity index $c$ on the jamming line, computed on the RS solution for $\s\geq 0$,
and on the fullRSB solution for $\s<0$.
The system is hypostatic for $\s>0$ and isostatic for $\s \leq 0$. 
}
\label{Contatti}
\end{figure}

Plugging the RS ansatz in Eq.~\eqref{eq:zperc}, and using the asymptotic scaling of $f(q_M,h)$ in Eq.~\eqref{eq:fasyunsat}, 
one gets the fraction of contacts in the UNSAT phase (at the replica symmetric level):
\beq
z=\int_{-\infty}^0 \frac{\de h}{\sqrt{2\pi}}e^{-\frac{(h+\s)^2}2} = \Th\left(  \frac{\s}{\sqrt{2}} \right)  \:.
\label{contactsRS}
\eeq
The isostaticity condition is $c = \a z =1$, and the isostaticity index $c$ is reported in Fig.~\ref{Contatti} along the jamming line.
For $\s>0$ the system is hypostatic on the jamming line, while it becomes isostatic at $\s=0$ \cite{FPUZ15}. 

\subsubsection{Stability of the RS solution}
\label{sec:RSstab}

After the RS solution, and the associated phase diagram, has been discussed, one should investigate the stability
of this solution towards replica symmetry breaking (RSB).
RSB can be associated to a continuous
de Almeida-Thouless (dAT) instability~\cite{MPV87}, or to the discontinuous appearance of a 
RSB solution, usually called a ``Random First Order Transition'' (RFOT)~\cite{CC05}. In this section we concentrate on
the first mechanism, which is the relevant one for the transition in
the UNSAT region and in the SAT region at moderately negative values of
$\sigma$. 

The dAT continuous instability of the RS solution can be discussed by
computing the eigenvalues of the
Hessian matrix, defined as~\cite{MPV87}
\beq
H_{ab;cd}=\left.\frac{\de^2 S[Q]}{\de q_{ab}\de q_{cd}}\right|_{q_{a\neq b}=q_M}
\eeq
on the RS saddle point solution $q_{a\neq b}=q_M$,
where $q_M$ satisfies the RS saddle point equation.
The continuous breaking of replica symmetry is associated to the vanishing of an eigenvalue of the
Hessian matrix.

Here,
instead of computing explicitly the eigenvalues of $H_{ab;cd}$, we follow an
equivalent procedure that makes use of the fullRSB equations derived in Sec.~\ref{sec:perc_def}.
Replica symmetry breaking means that the function $q(x)$ is not a constant; continuous RSB
means that $q(x)$ becomes non-constant in a continuous way, and therefore close to the instability $q(x)$ is very close
to a constant. Usually, the deviation from a constant is localized around a particular point $x$.
We thus assume that there is a single value of $x$ where $\dot q(x)$ is continuously 
becoming different from zero. 
Slightly in the unstable phase, around this point $x$,
Eq.~\eqref{eq:qx2} holds. Upon approaching the instability point from the unstable phase, the function $q(x)$ tends to a constant
and Eq.~\eqref{eq:qx2} reduces to its expression computed on the RS solution:
\beq\label{eq:dAT}
\frac{1}{(1-q_M)^2}=\a \int_{-\infty}^\infty \de h \g_{q_M}(h+\s)\left[\frac{\de^2}{\de h^2}\ln \g_{1-q_M}\star e^{-\b v(h)}\right]^2\:.
\eeq
This expression, computed on the value of $q_M$ that satisfies the RS saddle point equation,
 gives a condition on $(\a,\s)$ that must be satisfied at the dAT instability, hence defining the instability 
line $\a_{\mathrm{dAT}}(\s)$ in the phase diagram.
Note that computing Eq.~\eqref{breaking_point} on this line gives the ``breaking point'', i.e. the value of $x$ at which the instability occurs.

In the SAT phase ($T\to 0$ at finite $q_M$), Eq.~\eqref{eq:dAT} reduces to
\beq
\frac{1}{(1-q_M)^2}=\a \int_{-\infty}^\infty \de h \g_{q_M}(h+\s)\left[\frac{\de^2}{\de h^2}\ln \Th\left(\frac{h}{\sqrt{2(1-q_M)}}\right)\right]^2\:.
\eeq
In the UNSAT phase ($T\to 0$ with $q_M = 1-\chi T$), it instead becomes
\beq
\frac{1}{\chi^2}=\frac{\a}{(1+\chi)^2}\int_{-\infty}^\s \frac{\de z}{\sqrt{2\pi}}e^{-z^2/2} \ ,
\label{Replicon_unsat}
\eeq
and using Eq.~(\ref{RS_unsat}) that gives the value of $\chi$, we get that the transition line corresponds to $\s=0$, $\forall \a > \a_{\mathrm{J}}(\s=0)=2$.
Thus, there are two dAT transition lines where replica symmetry spontaneously breaks, one in the SAT and one in the UNSAT phase; they
are reported in Fig.~\ref{PD}.
An important conclusion of this study, which is consistent with Ref.~\cite{DG88}, is that in the whole region $\s>0$ the RS solution is stable; in particular,
in that region the jamming transition happens in the RS phase and the system is hypostatic at jamming.

\subsection{The nature of the RSB phase} 
\label{sec_nature}

We have seen in Sec.~\ref{sec:RSstab} that replica symmetry must be spontaneously broken in the region delimited by the dAT instability, reported in the phase
diagram of Fig.~\ref{PD}.
We now characterize the nature of the RSB transition and of the broken symmetry phase.
We follow a recipe based on experience with this kind of transitions, along the following logical steps:
\begin{enumerate}
\item
The first step is to determine the point at which the RS solution is unstable; this is the dAT line $\a_{\mathrm{dAT}}(\s)$ determined in Sec.~\ref{sec:RSstab}.
At the dAT line, the function $q(x)$ is not constant anymore.
As mentioned in Sec.~\ref{sec:RSstab}, the breaking point $m$ of the RSB solution, 
i.e. the point $x=m$ where $\dot q(x)$ becomes different from zero,
can be computed plugging the RS ansatz into Eq.~(\ref{breaking_point}). 
\item
The next step is to evaluate whether $m<1$ or $m>1$. In fact, because the function $q(x)$ is defined for $x\in[0,1]$, a consistent RSB solution
requires $m<1$. If this is not the case, then the dAT line cannot be a transition line, it must be preceded by a discontinuous transition of the RFOT kind. In fact
one can show that the case $m=1$ separates the two regimes. When $m=1$, the dAT instability splits into two RFOT-like transition lines: a ``dynamical transition'' where a
1RSB solution appears discontinuously but the free energy remains analytical, and a Kauzmann (or condensation) transition which corresponds to a true phase transition
to a spin glass phase~\cite{CC05,KMRSZ07}. We will discuss further this situation in Sec.~\ref{sec:1RSB}.
\item
If $m<1$, the dAT instability corresponds to a true continuous phase transition between a ``paramagnetic'' RS and a ``spin glass'' RSB phase. 
However, the dAT
instability can give rise either to a fullRSB phase, 
with a continuous $q(x)$ for $x\in[x_m,x_M]$,
or to a 1RSB phase, with $x_m = x_M = m$ and $q(x)= q_m$ for $0<x<m$ and $q(x)= q_M$ for $m<x<1$.
To determine which one is the case,
the final step is to investigate the
value of $\dot q(m)$. In fact, a fullRSB solution requires $\dot q(m)>0$, because $q(x)$ must be an increasing function\footnote{See \cite{Virasoro} for an attempt of interpreting
  decreasing solutions in the replica formalism.} of $x$. 
If this is not the case, then the transition is a continuous transition to a 1RSB solution.
To compute $\dot q(m)$, we derive Eq.~\eqref{breaking_point}, expressed as a function of $x$, 
with respect to $x$. We get
\beq\label{eq:treder}
 \begin{split}
\dot q(x) &=\left\{\l^3(x)\left[\frac \a2 \int
    \de h P(x,h) A(x,h) - \frac{3 x^2}{\l^4(x)}\right]\right\}^{-1} \ , \\
A(x,h)&= f''''(x,h)^2-12 x
f''(x,h) f'''(x,h)^2+6x^2
f''(x,h)^4 \ .
\end{split}\eeq
 By evaluating Eq.~\eqref{eq:treder} on the RS solution, with $x=m$, we obtain
 the desired result.

\end{enumerate}
These steps can be performed for both dAT instabilities, in the SAT and UNSAT phases, leading to a full characterization
of the RSB transition.

\subsubsection{The SAT phase}

In the SAT phase, the breaking point is
\beq
\begin{split}
m=\frac{1-q_M}{2}\frac{\int \de h \g_{q_M}(h+\s) f'''_{\mathrm{SAT}}(q_M, h)^2}{\int  \de h \g_{q_M}(h+\s) f''_{\mathrm{SAT}}(q_M, h)^2\left[1-(1-q_M)f''_{\mathrm{SAT}}(q_M, h)\right]} \ ,\\
\end{split}
\eeq 
where $f_{\rm SAT}(q_M,h)$ is defined in Eq.~\eqref{eq:fSAT}, and $q_M$ is the solution of the saddle point equation
(\ref{RS_sp}) computed at $\a=\a_{\mathrm{dAT}}(\s)$. 
A numerical evaluation of $m$ as a function of $\s<0$, on the dAT line in the SAT phase,
shows
that close enough to $\s=0$ one has $m<1$, while $m$ increases upon decreasing $\s$ towards more negative values.
At $\s=\s_{\mathrm{RFOT}}$ one has $m=1$. Beyond that point,
for $\s<\s_{\mathrm{RFOT}}$, one has $m>1$ and the transition becomes discontinuous, in the RFOT universality class.
In order to compute
the properties of the phase diagram for $\s<\s_{\mathrm{RFOT}}$ we
need to study the 1RSB solution more carefully, see Sec.~\ref{sec:1RSB}.

Furthermore, the analysis of $\dot q(m)$ on the same transition line shows that 
close enough to $\sigma=0$, for $\s>\s_{\mathrm{1RSB}}$, one has $\dot q(m)>0$ and the dAT instability leads
to a fullRSB phase, while
for $\s \in [\s_{\mathrm{RFOT}},\s_{\mathrm{1RSB}}]$ one has $\dot q(m)<0$ and
the instability leads to a continuous transition from a RS phase
towards a 1RSB phase. These results are shown in the phase diagram of Fig.~\ref{PD}.

\subsubsection{The UNSAT phase}

Defining 
\beq\label{eq:FFdef}
\FF(x)\equiv \ln \int_{-\infty}^\infty \frac{\de z}{\sqrt{2\pi \chi}} \exp\left[-\frac{(x-z)^2}{2\chi}-\frac 12 z^2 \th(-z)\right]\:,
\eeq
in the zero temperature limit in the UNSAT phase the breaking point is
\beq
m\simeq \sqrt{T}\chi (1+\chi)^3 \int \frac{\de x}{\sqrt{2\pi}} \FF'''(x)^2\equiv \hat m \sqrt{T}\:.
\label{m_UNSAT}
\eeq 
The breaking point thus tends to zero proportionally to $\sqrt{T}$, and therefore
in the UNSAT phase the dAT instability always leads to a consistent continuous phase transition
(a discontinuous transition is never present in this case).
We remark that the scaling of the breaking point along the dAT line, $m\propto \sqrt{T}$, 
is different from the one observed in the Sherrington-Kirkpatrick model in the zero temperature limit,
where instead the breaking point remains finite. 
The origin of this difference will be clarified in Sec.~\ref{sec:SATUNSAT}.

\begin{figure}
\centering
\includegraphics[scale=0.4]{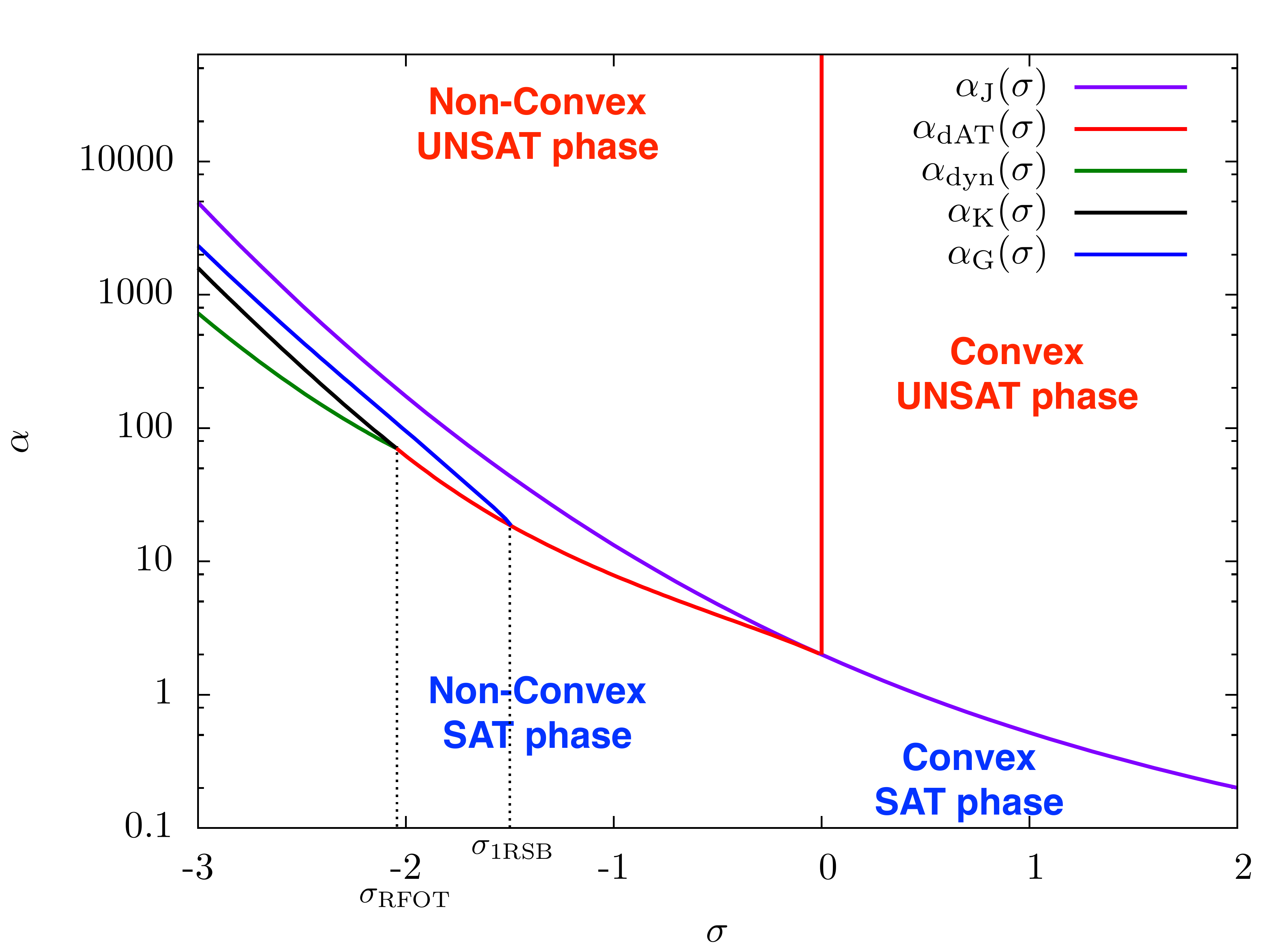}
\caption{The zero temperature phase diagram of the perceptron. $\a_J(\sigma)$ is the jamming transition (or SAT/UNSAT threshold) within the replica symmetric approximation. The replica symmetric solution is stable for all $\s>0$. Thus the jamming transition line is exact only for $\s>0$.
$\a_{\mathrm{dAT}}(\s)$ are the lines where replica symmetry breaks. In the non-convex region and for $\s\in[\s_{\mathrm{1RSB}},0]$, $\a_{\mathrm{dAT}}$ is a transition from a replica symmetric phase to a continuous fullRSB one.
For $\s\in [\s_{\mathrm{RFOT}},\s_{\mathrm{1RSB}}]$ the dAT line is a continuous transition from a replica symmetric phase to a stable 1RSB phase.
For $\s<\s_{\mathrm{RFOT}}$ a RFOT
type phenomenology is observed. Keeping $\s$ fixed and increasing $\a$, one has first a
dynamical transition for $\a=\a_{\mathrm{dyn}}(\s)$, then
a Kauzmann transition for $\a=\a_{\mathrm{K}}(\s)$, and finally a Gardner transition for $\a=\a_{\mathrm{G}}(\s)$.
In the UNSAT phase, the instability line $\a_{\mathrm{dAT}}(\s)$ represents the transition from a replica symmetric phase to a continuous fullRSB one for all $\a>2$.
Therefore, fullRSB occurs in the region delimited by the $\a_{\mathrm{dAT}}(\s)$ and $\a_{\mathrm{G}}(\s)$ lines, which contains the whole jamming line for $\s<0$.
}
\label{PD}
\end{figure}

Moreover, the zero
temperature limit of the slope $\dot q(m)$ at the breaking point, in the UNSAT phase, is given by
\beq \dot q(m)\simeq \left[\frac{\a \chi^3}{2}\frac{1}{\sqrt{T}} \int
  \frac{\de x}{\sqrt{2\pi}} \FF''''(x)^2\right]^{-1}\geq 0  \ ,
\eeq
which is of order $\sqrt{T}$ and always positive. This
implies that the dAT instability line in the UNSAT phase is a transition
from a RS solution to a fullRSB one~\cite{FPUZ15}, see Fig.~\ref{PD}. 
We will see in Sec.~\ref{sec:SATUNSAT} that the
scaling with $T$ of both $m$ and $\dot q(m)$ is in perfect agreement with the complete scaling
form of $q(x)$ in the UNSAT phase.

\subsection{The 1RSB free-energy in the SAT phase}
\label{sec:1RSB}

In Sec.~\ref{sec_nature} we have shown that for $\s<\s_{\mathrm{RFOT}}$ the solution for $q(x)$ in the SAT phase becomes of the RFOT type 
(i.e. a discontinuous 1RSB solution)~\cite{CC05}. In order to characterize this part of the phase diagram we need to consider explicitly a general 1RSB ansatz.
The 1RSB free energy is characterized by three numbers $q_0=q_m,\;
q_1=q_M$ and $m$, the function $q(x)$ being equal to $q_m$ for $x\in [0,m]$ and to $q_M$ for $x\in [m,1]$
(see Appendix~\ref{sec:KRSB} for a general $K$RSB solution).
At zero temperature, in the SAT phase where both $q_0<1$ and $q_1<1$, 
the function $-\b {\rm f}[x(q)]$ has a finite limit, which coincides with the 1RSB entropy, given by
\begin{eqnarray}
\nonumber
 s_{1RSB}(q_1,q_0,m) &=& \frac{m-1}{2m}
 \log(1-q_1)+\frac1{2m}\log(1+m(q_1-q_0)-q_1) \\ &+& \frac{q_0}{2(1+m(q_1-q_0)-q_1)}+
  \frac \alpha m \int \frac{\de t}{\sqrt{2\pi q_0}}\; e^{-\frac{t^2}{2q_0}} \log\left[
\gamma_{q_1-q_0}
\star
\ \Th^m \left( \frac {t-\sigma}{
 \sqrt{2(1-q_1)}}\right) \right] \ .  
  \label{eq:11}
  \end{eqnarray}
The variational equations always admit the RS solution
$q_1=q_0$. For any given $\sigma$ this is the unique solution for
sufficiently small $\alpha$. 
For fixed $\s<\s_{\mathrm{RFOT}}$ and upon increasing $\a$, there is a point, called the dynamical transition $\a_{\rm dyn}(\s)$, where a
non-trivial solution with $q_1\ne q_0$ appears for $m\to 1$. We do not discuss in details the properties of this transition, for which
we refer to~\cite{Mo95,KMRSZ07,CC05}. 
In short, at $\a_{\rm dyn}(\s)$ the Boltzmann-Gibbs measure on the space of solutions splits into an exponential number of clusters of solutions~\cite{KMRSZ07}.
These clusters can be classified according to their internal entropy that measures their typical size. 
At the dynamical transition point, the clusters that dominate the Boltzmann-Gibbs measure are the 
most numerous ones, meaning the ones with higher complexity (or configurational entropy). It is important to note that the free energy remains
analytic upon crossing
$\a_{\rm dyn}(\s)$, which is thus not a thermodynamic phase transition~\cite{KMRSZ07,CC05}.
Increasing $\a$, the complexity of the relevant clusters decreases and vanishes 
at the Kauzmann (or condensation) transition $\a=\a_{\textrm{K}}(\s)$~\cite{Mo95,KMRSZ07,CC05}.
At this point, the clusters that dominate the Boltzmann-Gibbs measure are no longer exponentially many, and the system undergoes a thermodynamic
phase transition. Note that this two-transition scenario, also called RFOT, is at the basis of the mean field theory of glasses
(see~\cite{BB11,Ca09,PZ10,WL12,KT15} for reviews).

To compute $\a_{\rm dyn}(\s)$ and $\a_{\rm K}(\s)$, 
we are therefore led to consider the 1RSB entropy for 
$m\approx 1$, where we can write to the leading order
\begin{eqnarray}
  \label{eq:12}
s_{1RSB}(q_1,q_0,m)=s_{RS}(q_0)+(m-1)\d s_{RSB}(q_0,q_1) \ ,
\end{eqnarray}
with $s_{RS}$ given in Eq.~\eqref{s_RS}, and
\beq \begin{split}
\label{frsb}
\d s_{RSB}(q_0,q_1)&= -s_{RS}(q_0) - \frac{q_0^2}{2(1-q_0)^2}  + \frac{q_1(1-2q_0)}{2(1-q_0)^2}
 +\frac12 \log (1-q_1) 
\\ & +\alpha \int{ \frac{\de t}{\sqrt{2\pi q_0}}}\;
 e^{-\frac{t^2}{2q_0}}\int  \frac{\de u}{\sqrt{2\pi (q_1-q_0)}}\; 
e^{-\frac{u^2}{2(q_1-q_0)} }\ \frac{\Th\left( \frac{u+t-\sigma} {\sqrt{2(1-q_1)}}\right)}{\Th\left(
     \frac{t-\sigma}{\sqrt{2(1-q_0)}}\right) } \log \Th\left( \frac{u+t-\sigma} {\sqrt{2(1-q_1)}}\right) \:. 
 \end{split}\eeq
 When $m\to 1$, 
 $q_0$ can be obtained by setting to zero the derivative of Eq.~(\ref{s_RS}), which gives the saddle point equation~(\ref{RS_sp}).
 The dynamical transition point $\a_{\rm dyn}(\s)$ corresponds to the lowest value of $\a$ where the variational equation for $q_1$, obtained by setting the derivative of
 Eq.~(\ref{frsb}) to zero, admits a solution $q_1>q_0$.
At the Kauzmann transition point $\a_{\rm K}(\s)$, the breaking point $m$, obtained by setting the derivative with respect to $m$ 
of Eq.~(\ref{eq:11}) to zero, first becomes smaller
than one. 
Upon further increasing $\a$, it can be shown that the 1RSB solution becomes unstable and undergoes a Gardner transition
towards a fullRSB phase~\cite{Ga85}.
The equation that controls the Gardner instability of the 1RSB solution can also be obtained using the fullRSB equations, similarly to what has 
been done in Sec.~\ref{sec:RSstab} for the RS solution. In the following we assume that the continuous part of $q(x)$ appears in correspondence
to the value $q_1$, as it usually happens at a Gardner transition~\cite{Ga85} (the stability towards a continuous breaking at $q_0$ can be discussed
along similar lines).
If $q_1$, $q_0$ and $m$ satisfy the saddle point conditions, the instability point $\a_{\rm G}(\s)$ of the 1RSB transition is given by the following condition:
\beq
\frac{1}{\l_{1RSB}^2(1)}=\a \int \de h\, P_{1RSB}(1,h) \, f''_{1RSB}(1,h)^2 \ ,
\eeq
where
\beq
\begin{split}
\l_{1RSB}(1)&=1-q_1 \ , \\
f_{1RSB}(1,h)&=\log \Th\left[\frac{h}{\sqrt{2(1-q_1)}}\right] \ , \\
P_{1RSB}(1,h)&=\Th^m\left[\frac{h}{\sqrt{2(1-q_1)}}\right]\g_{q_1-q_0}\star\left[\g_{q_0}(h+\s)e^{-mf_{1RSB}(m,h)}\right] \ , \\
f_{1RSB}(m,h)&=\frac{1}{m}\log \left[\g_{q_1-q_0}\star \Th^m\left[\frac{h}{\sqrt{2(1-q_1)}}\right]\right] \ .
\end{split}
\eeq
The dynamical, the Kauzmann and Gardner transition lines are plotted in Fig.~\ref{PD}.

\section{The SAT-UNSAT transition and its critical properties}
\label{sec:SATUNSAT}

Having established the phase diagram of the zero temperature random perceptron model (Fig.~\ref{PD}), 
we discuss here the critical properties of the jamming (or SAT-UNSAT) transition line.
The jamming line at $\s>0$ has been discussed already in \cite{DG88, FP15} and it is known to be non-critical,
in the sense that the system is hypostatic (see Fig.~\ref{Contatti}), which according to the analysis of~\cite{Wy12}
does not lead to marginal stability.
In this section we want to describe the jamming transition for $\s<0$,
which falls in the fullRSB phase (Fig.~\ref{PD}): in this case, 
the system is isostatic at jamming and a 
non-trivial critical behavior appears~\cite{FP15}.
The jamming point can be approached both from the SAT and UNSAT phase.
From the SAT (unjammed) phase, upon approaching the transition
the volume of the space of solutions shrinks to zero and the self-overlap in a cluster of solutions
is asymptotically close to one. 
From the UNSAT (jammed) phase, the energy goes to zero upon approaching the transition.
In both cases,
universality naturally emerges with a set of nontrivial critical exponents that characterize the scaling
of physical quantities on both sides of the transition.
We did not attempt to solve numerically the fullRSB equations (see~\cite{CKPUZ14} for a numerical study
of the equations in the hard sphere case); in fact, the values of the critical exponents can be extracted analytically
via a scaling analysis of the equations, that we discuss in the rest of this section.

\subsection{Asymptotes of $f(q,h)$ and $P(q,h)$}

For later use let us discuss the asymptotic behavior of $f(q,h)$ and
$P(q,h)$ for $h\to \pm\infty$ which are generically valid, in
particular in the scaling solutions of our interest.  For $h\to
\pm\infty$ the boundary condition for $f(q,h)$ reduces to $f(q_M,h\to
\infty)=0$ and $f(q_M,h\to
-\infty)=-\frac{h^2}{2}\frac{\beta}{1+\beta(1-q_M)}$. Using Eq.~(\ref{eqf})
one readily finds 
\beq \label{eq:fqhasy}
f(q,h\to\io) = 0 \ , \hskip30pt f(q,h\to-\io)
\sim -\frac{h^2}{2} \frac{\b}{1+\b \l(q)} \ .  
\eeq 
For $h\to\io$,
Eq.~\eqref{eq:qP} becomes simply $\dot P = P''/2$, which has a unique
solution compatible with the boundary condition 
\beq\label{eq:P+}
P(q,h\to \io) = \g_q(h+\s) \ .  
\eeq 
For $h\to -\io$ the equation is
slightly more complicated; but still, a Gaussian form of
the kind $P(q,h) \sim \sqrt{D(q)} e^{ - D(q) h^2}$ satisfies
Eq. (\ref{eq:qP}) for large negative $h$, where $D(q)$ verifies
\begin{eqnarray}
\label{eq:P-}
&&\dot D(q) = -2 D(q)^2 + 2 D(q) \frac{\b x(q)}{1+\b\l(q)} \ ,\nonumber\\
&& D(q_m)=\frac{1}{2q_m} \ .
\end{eqnarray}

\subsection{Approaching jamming from the SAT phase}
\label{sec:jamSAT}

In the SAT phase we can take the limit $T=1/\b\to 0$ by simply
replacing $e^{-\b v(h) } \to \th(h)$. The volume of the space of solutions is finite and
the resulting equations are well defined and give a value
$q_M<1$. This is due to the fact that zero-energy configurations are
not isolated: they form clusters where two typical solutions have overlap $q_M <
1$. Solutions that belong to different clusters have overlap $q<q_M$ whose statistical
properties are described by the function $x(q)$: it represents the (average)
probability that two configurations have an overlap smaller than $q$ \cite{MPV87}. 
In the jamming limit,
 the cluster volumes go to zero and therefore the self-overlap of a cluster, $q_M$, goes to one. 

\subsubsection{Scaling form of the solution close to jamming}

Close to jamming the fullRSB equations develop a scaling regime, as can be deduced from a numerical analysis~\cite{CKPUZ14}
(see Sec.~\ref{sec:num_sol} and Appendix~\ref{sec:KRSB} for details on how to solve the fullRSB equations numerically).
In the scaling regime, it is convenient to make the following change of variables:
\beq\label{eq:scalSAT}
\begin{split}
y(q) &= \ee^{-1} x(q)  \ , \\
\wh f(q,h) &= \ee f(q,h) \ , \\
\wh \l(q) &= \ee^{-1} \l(q)  \ , \\
\end{split}
\eeq
where $\ee$ is the linear distance from the jamming line (either in $\a$ or $\s$).
In addition we assume that $q_M = 1 - \ee^\k$ where $\kappa$ is an
exponent to be determined by the equations. 
In the jamming limit $\ee\to 0$, one has $y \in [0,1/\ee] \to [0,\io)$ and
$q \in [q_m, q_M] \to [q_m , 1]$.

We wish to show that a scaling solution exists in this limit, when $y$ is large and $q \sim 1$. It has the following form:
\beq
\label{eq_scaling_sol}
\begin{split}
& y(q) \sim y_J  (1 - q)^{-1/\k}  \\
&P(q,h)  \sim  \begin{cases}
(1-q)^{(1-\k)/\k} p_-[ h (1-q)^{(1-\k)/\k} ] & \text{for } h \sim -(1-q)^{(\k-1)/\k} \\
(1-q)^{-a/\k} p_0\left( \frac{h}{\sqrt{1-q}} \right) & \text{for } |h| \sim \sqrt{1-q} \\
p_+(h) & \text{for } h \gg \sqrt{1-q}
\end{cases} \\
& m(q,h) = \wh \lambda(q) \wh  f'(q,h)
= -\sqrt{1-q} \MM \left[ \frac{h}{\sqrt{1-q}} \right] \ , \hskip20pt \MM(t\to\io)=0 \ , \hskip20pt \MM(t\to-\io) = t \ . 
\end{split}
\eeq
While the functions $p_-$ and $p_+$  and the equations that they
verify are peculiar of the
perceptron model and depend on the parameters $\sigma$ and $\alpha$,
on the jamming line the function $p_0$ as well as the exponents $a$
and $\kappa$ are  universal (i.e. independent of
the precise location on the jamming line).  In Sec.~\ref{sec:Jproof}
we show that the universal equations
determining $p_0$ and the critical exponents also coincide with the ones obtained for the jamming transition of hard spheres in high dimension.
Note that for finite $\ee$, the scaling solution (\ref{eq_scaling_sol}) is cutoff when $q \sim q_M = 1 - \ee^\k$ and $y\sim y_M = Y \ee^{-1}$.
Finally, note that while we will be able to prove the existence of the scaling solution, and compute the values of $a$ and $\k$, we will not
be able to prove that $\ee$ is proportional to the distance from the jamming line:
at present, this follows from the numerical solution of the equations~\cite{CKPUZ14}.

\subsubsection{Proof of the scaling form}
\label{sec:Jproof}

The scaling analysis is carried out along the lines of~\cite{CKPUZ13,CKPUZ14,FP15}. 
\medskip

\noindent
{\bf Scaling of $f(q,h)$ --}
The function
$m(q,h)$
introduced in Eq.~\eqref{eq_scaling_sol} satisfies the equation
\beq\label{eq:eqM}
\dot m(q,h) 
= -\frac12 m''(q,h) - \frac{y(q)}{\wh\l(q)} m(q,h) [ 1 + m'(q,h) ]  \ , \qquad m(q,h\to -\infty)\simeq -h
\ , \qquad
 m(q,h\to \infty)\simeq 0
\ ,
\eeq
where the differential equation comes from Eq.~\eqref{eqf} and the asymptotes from Eq.~\eqref{eq:fqhasy}.
Let us inspect the value of $\frac{y(q)}{\wh\l(q)}$ for $\epsilon\to
0$ and $q\to 1$, according to Eq.~\eqref{eq_scaling_sol}:
\beq\label{eq:ysul}
\frac{y(q)}{\wh\l(q)} = \frac{y_J (1-q)^{-1/\k}}{ \ee^{\k-1} [1-y_J \frac{\k}{\k-1}] + y_J \frac{\k}{\k-1} (1-q)^{1-1/\k}}
\ .
\eeq
This expression has a crossover for $1 - q \sim \ee^\k$, when the two terms in the denominator are of the same order.
The scaling form is obtained for $1 - q \gg \ee^\k$, in which case we have 
\beq\label{eq:ysul2}
\frac{y(q)}{\wh\l(q)} = 
\frac{\k-1}\k \frac1{1-q}
\ .
\eeq
Plugging this result and the scaling form (\ref{eq_scaling_sol}) in Eq.~\eqref{eq:eqM}, we obtain
a scaling equation for the function $\MM(t)$:
\beq\label{eq:eqMM}
\MM(t) - t \MM'(t)
=  \MM''(t)
 +2 \frac{\k-1}\k  \MM(t)
  [ 1 - \MM'(t) ] 
\ ,
\hskip20pt \MM(t\to\io)=0 \ , \hskip20pt \MM (t\to-\io) = t \ . 
\eeq
This non-linear equation, with boundary conditions at $t\to\pm\io$, admits a
unique solution for each value of $\k$. 

\medskip

\noindent
{\bf Scaling of $P(q,h)$ --} The existence of the functions $p_-$ and
$p_+$ and the corresponding scaling variables can be obtained by the
analysis of the equation for $P$ at large negative and positive
arguments, respectively.
From Eq.~\eqref{eq:P+} it follows that for $h\to\io$, $P(q,h)$ remains a finite Gaussian,
$P(q,h\to\io) \sim \g_q(h+\s)$. For finite $h>0$, the Gaussian
will be deformed to a finite function $p_+(h)$ as it appears in Eq.~\eqref{eq_scaling_sol}. 
For $h\to -\io$, on the other hand, we have from Eq.~\eqref{eq:P-} that for $\b\to\io$ and in the scaling regime
of Eq.~\eqref{eq:ysul2}:
\beq
\dot D(q) = -2 D(q)^2 + 2 D(q) \frac{y(q)}{\wh\l(q)} =  -2 D(q)^2 + 2 D(q)  \frac{\k-1}\k \frac1{1-q} \ .
\eeq
If $\k<2$ (which, we will see, is the case), this equation admits a
scaling solution  $D(q) \sim D_J \, (1-q)^{-2(\k-1)/\k}$ where the term $2
D(q)^2$ is negligible.  One concludes therefore that in the scaling regime,
$P(q,h\to-\io) \sim \sqrt{D(q)}e^{-D(q) h^2}$ has the form
\beq
P(q,h) \sim (1-q)^{(1-\k)/\k} p_-[ h (1-q)^{(1-\k)/\k} ].
\eeq
which has been proven asymptotically for $h\to-\io$ but
can be extended to the whole regime
where $q\sim 1$ and $|h | \sim (1-q)^{(\k-1)/\k}$, as in Eq.~\eqref{eq_scaling_sol}. 
Note that even if it has a scaling form, the function $p_-$ is {\it
  not} uniquely determined by the scaling regime, and remains
non-universal, see~\cite{CKPUZ13}. 

The ``matching'' regime with $p_0(t)$ must be introduced in Eq.~\eqref{eq_scaling_sol} 
to smoothly match the two regimes for negative and positive $h$. Here
is where universality appears.
Matching $p_0$ and $p_+$ requires that
\beq\label{eq:73}
p_+(t\to 0^+) \sim t^{-\g} \ ,
\qquad
p_0(t \to \io) \sim t^{-\g} \ ,
\qquad
\g =  \frac{2a}{\k} \ .
\eeq
Matching $p_-$ and $p_0$ requires that 
\beq\label{eq:74}
p_-(t\to 0^-) \sim |t|^\th \ ,
\qquad
p_0(t \to -\io) \sim |t|^{\th} \ , 
\qquad 
\th = \frac{1-\k+a}{\k/2-1} \ .
\eeq
The scaling variable $t = h/\sqrt{1-q}$ appearing
in the matching regime is naturally the same as for $f(q,h)$. This is in fact the only choice that leads,
once plugged in Eq.~\eqref{eq:qP},
to a non-trivial equation for $p_0(t)$, namely:
\beq\label{eq:eqp1}
\frac{a}{\k} p_0(t) + \frac12 t \, p_0'(t) = \frac12 p_0''(t)
+\frac{\k-1}\k  ( p_0(t) \MM(t) )' \ ,
\hskip15pt p_0(t\to\io)=t^{-2a/\k} \ , \hskip15pt p_0(t\to-\io) = |t|^{ (1-\k+a)/(\k/2-1) } \ .
\eeq
Recall that in Eq.~\eqref{eq:eqp1}, $\MM(t)$ depends on $\k$; 
it turns out that Eq.~\eqref{eq:eqp1} also admits a unique solution for $p_0(t)$ satisfying the correct asymptotic
conditions, {\it but only for a given choice of $a=a(\k)$}. For a given $\k$, Eqs.~\eqref{eq:eqMM} and \eqref{eq:eqp1} thus determine
$\MM(t)$, $p_0(t)$ and $a$.

\medskip

\noindent
{\bf Determination of the exponent $\k$ --} The exponent $\k$ can be fixed using Eq.~\eqref{breaking_point} which 
can be equivalently written as
\beq\label{eq:margsc}
\frac{y(q)}{\wh \l(q)} = \frac{1}2 
\frac
{ \int \de h P(q,h) m''(q,h)^2  } 
{ \int \de h P(q,h) m'(q,h)^2 [1 +  m'(q,h)] } 
\ .
\eeq
In the scaling regime, Eq.~\eqref{eq:ysul2} gives the left hand side. In the right hand side, we can note that
$m''(q,h)$ and $m'(q,h)^2 [1 +  m'(q,h)]$ both vanish outside the scaling regime, because of the asymptotic
behavior of $m(q,h)$. Therefore, the right hand side only receives contribution from the regime $h \sim \sqrt{1-q}$.
We obtain
\beq\label{eq:kappa}
 \frac{\k-1}\k 
= \frac{1}2 
\frac
{ \int \de t \, p_0(t) \MM''(t)^2  } 
{ \int \de t \, p_0(t) \MM'(t)^2 [1 +  \MM'(t)] } 
\ .
\eeq
Because both $\MM(t)$ and $p_0(t)$ depend on $\k$, this is an equation for $\k$.
The numerical solution of the system of Eqs.~\eqref{eq:eqMM}, \eqref{eq:eqp1} and \eqref{eq:kappa},  gives
$\k=1.41574\ldots$ while within the numerical precision one finds the relation~\cite{CKPUZ13}
\beq\label{eq:ak}
a=1 - \frac{\k}2 \qquad
\Rightarrow \qquad
\g = \frac{2-\k}\k \ , \ \  \th = \frac{3\k-4}{2-\k} 
\text{ and } \g = \frac1{2+\th}
\ .
\eeq
The importance of these ``scaling'' relations will be further discussed in Sec.~\ref{sec:scaling} and Sec.~\ref{sec:gammatheta}.

\subsection{Zero temperature fullRSB solution in the UNSAT phase} 
\label{sec:scaling_UNSAT}
  
 We now turn to the analysis of the approach to the jamming transition from the UNSAT phase. Before doing that, however,
 we need to study the behavior of the fullRSB solution for $T\to 0$ in the UNSAT phase.
In Sec.~\ref{sec_nature} we have shown that in the UNSAT phase, close to the dAT instability line $(\s=0, \a>2)$ of the RS solution,
the solution $q(x)$ has the following properties:
\begin{itemize}
\item{The Edwards-Anderson order parameter is $q_M=1-\chi T$}
\item{The breaking point at the instability transition line is $m=\hat m \sqrt{T}$}
\item{The slope of $q(x)$ at the breaking point on the instability line is $\dot q(m)\sim \sqrt{T}$}
\end{itemize}
Additionally, here we want to show that, like in the Sherrington-Kirkpatrick model~\cite{PT80,MPV87}:
\begin{itemize}
\item{$P(q,h)$ is smooth and regular at zero temperature}
\item{The function $\beta x(q)$ admits a finite zero temperature limit}
\end{itemize}
For small temperature in the UNSAT phase, the initial condition for $f(q_M,h)$ is given by  
(Appendix~\ref{App_Asy}):
\beq
 f(q_M,h) = \begin{cases}
-\frac{\b h^2}{2(1+\chi)}\th(-h) & \text{for } |h|\gg \sqrt{T} \ , \\
\FF(h/\sqrt T) & \text{for } |h| \sim \sqrt T \:,
\end{cases}
\eeq
with $\FF(x)$ defined in Eq.~\eqref{eq:FFdef}.
In the fullRSB region, as in the RS case, we define $q_M=1-\chi T$ and we introduce
\beq  \label{eq:9}
y(q) = \frac{\b}\c x(q) \ , \qquad 
\wh f(q,h) = T \c f(q,h) \ ,\qquad
  \wh \l(q) = \frac{\b}\c \l(q) = 1 + \int_q^{1} \de p\; y(p)\:.
\eeq
As in Sec.~\ref{sec:Jproof}, we introduce $m(q,h) = \wh \lambda(q) \wh f'(q,h)= \lambda(q) f'(q,h)$,
which satisfies Eq.~(\ref{eq:eqM}) with the modification $m(q,h\to-\io) = - h \chi \wh\l(q)/(1+\chi \wh\l(q))$ and
initial condition
\beq\label{eq:qfS}
m(1,h) = - \frac{h \c}{1+\c} \th(-h)  \ . 
\eeq
The equation for $P(q,h)$ is still Eq.~\eqref{eq:qP}.
Finally,
Eq.~\eqref{eq:qx2} becomes
\beq\label{eq:chiS}
\frac{1}{\c^2} = \a \frac{\int \de h P(1,h) \th(-h)}{(1+\c)^2} \ ,
\eeq
and Eq.~\eqref{breaking_point} becomes identical to Eq.~\eqref{eq:margsc}.
The distribution of gaps, given in Eq.~\eqref{eq:rhoh}, becomes
\beq
\label{gab_distr}
\r(h) = \begin{cases}
P(1,h (1+\c) ) (1+\c) & \text{for } h< 0 \ , \\
P(1,h) & \text{for } h > 0 \ .
\end{cases}
\eeq
Because the breaking point $m = x(q_M) \propto \sqrt{T}$, one has
$y(q_M)\sim 1/\sqrt{T}$ and thus $y(q)$ extends up to infinity in the zero temperature limit. 
The scaling behavior of $y(q)$ for $q\to 1$ is expected to be \cite{PT80}:
\beq
\label{y_UNSAT}
y(q)\simeq \frac{y_\chi}{\sqrt{1-q}} \ .
\eeq
We now
check that this scaling is consistent.
First of all, this implies that $q(x)\sim 1- A T^2/x^2$ for some constant $A$. For $x=\hat m \sqrt{T}$, we get $q_M=1-\chi T$ which matches the behavior on the instability line.
Also, one has $\dot q(x) \propto T^2/x^3$ and then $\dot q(m) \propto \sqrt{T}$, as it is the case along the instability line.
The final check can be obtained from Eq.~(\ref{eq:margsc}).
We assume that the scaling behavior of $m(q,h)$ for $q\to 1$ is 
\beq  \label{eq:Tscal}
\begin{split}
m(q,h) \sim - \frac{\c}{1+\c} \sqrt{1-q} \MM\left( \frac{h}{\sqrt{1-q}}\right) 
\ ,
\hskip20pt \MM(t\to\io)=0 \ , \hskip20pt \MM(t\to-\io) = t \ , \\
\end{split}
\eeq
which agrees with the initial condition (\ref{eq:qfS}).
Plugging this ansatz inside Eq.~(\ref{eq:margsc}) we get
\beq\label{eq:ychi}
y(q)=\frac 12 \frac{1+\chi}{\sqrt{1-q}}\frac{P(1,0)\int_{-\infty}^\infty \de t (\MM''(t))^2}{\int_{-\infty}^0 \de hP(1,h)} 
\qquad
\Rightarrow
\qquad
y_\chi = \frac 12 (1+\chi) \frac{P(1,0)\int_{-\infty}^\infty \de t (\MM''(t))^2}{\int_{-\infty}^0 \de hP(1,h)} \ ,
\eeq
which confirms the consistency of Eq.~(\ref{y_UNSAT}) and provides an explicit expression of $y_\chi$.
The only point left to verify is that $P(1,0)$, and more generally
$P(1,h)$ 
are finite. First, we note that for $h\to \io$ we have $P(q,h) \to \g_q(h+\s)$, which suggests
that for $h\to\io$, $P(1,h)$ is finite and has smooth corrections in $q$.
Next, we can observe that Eq.~\eqref{eq:P-} becomes for $q\to 1$:
\beq
\dot D(q) = -2 D(q)^2 + 2 D(q) \left( \frac{y_\c}{\sqrt{1-q}} + \cdots \right)
\hskip20pt
\Rightarrow
\hskip20pt
D(q) = D_0 - 4 D_0 y_\c \sqrt{1-q} + \cdots
\eeq
which therefore indicates that for $h\to-\io$, $P(1,h)$ is finite and has corrections 
proportional to $\sqrt{1-q}$. 
Plugging $P(q,h)\sim  P(1, h) + \sqrt{1-q} \, \d P(h)$ in Eq.~\eqref{eq:qP}, 
and using Eq.~\eqref{eq:Tscal}, we obtain
\beq\label{eq:dPh}
\d P(h) = -2 y_\c \frac{\c}{1+\c} \lim_{q\to1} \frac{\de}{\de h} \left[ P(1,h) \sqrt{1-q} \MM\left( \frac{h}{\sqrt{1-q}} \right) \right] 
= -2 y_\c \frac{\c}{1+\c} \th(-h) [ h P'(1,h) + P(1,h) ] 
\ .
\eeq
which satisfies the condition $\int \de h \, \d P(h)=0$ and shows that
there are $\sqrt{1-q}$ corrections only in the region $h<0$.

\subsection{The jamming limit from the UNSAT phase}
\label{sec:jammingUNSAT}

The jamming limit from the UNSAT phase is obtained by considering the $T=0$ equations of Sec.~\ref{sec:scaling_UNSAT} 
in the limit $\chi\to\io$, as in the RS case.
This is because $q_M$ is finite in the SAT phase when $T=0$,  while in the UNSAT phase $q_M = 1 -\c T$ for $T\to 0$: 
matching the two regimes for $T\sim 0$ requires the divergence of $\chi$ at the jamming point.

We know that in the UNSAT phase for $q\to 1$ we have, from Eq.~\eqref{y_UNSAT} and \eqref{eq:9}:
\beq\label{eq:app65982}
\frac{ y(q)}{\wh \l(q)} \sim \frac{y_\c/\sqrt{1-q}}{ 1 + 2 y_\c \sqrt{1-q}} \ .
\eeq
If $\c$ is finite and $q\to 1$, then in the denominator the factor 1 dominates and $y/\wh\l \sim (1-q)^{-1/2}$: in this 
regime we recover the ``regular'' zero temperature solution of the UNSAT phase described in Sec.~\ref{sec:scaling_UNSAT}.
Conversely, for $\c\to \io$, the coefficient $y_\c \to \io$. In fact, from Eq.~\eqref{eq:ychi} we obtain $y_\chi  \propto \chi P(1,0)$
observing that $\MM(t)$ is finite and that $\int_{-\io}^0 \de h P(1,h)$ must remain finite because $P(1,h)$ is normalized to 1.
In this case,
 for all $q<1$, at large enough $\chi$ the second term in the denominator of Eq.~\eqref{eq:app65982} dominates
 and one has 
 \beq
 \frac{y(q)}{\wh\l(q)} \sim \frac1{2(1-q)} \ ,
 \hskip20pt
\text{for } q \sim 1 
\text{ and }  y_\c \sqrt{1-q} \sim \chi P(1,0) \sqrt{1-q}  \gg 1 \ . 
 \eeq
In this regime, we must have a different scaling solution in which $y/\wh\l \propto 1/(1-q)$: but this is exactly the jamming scaling solution
that was already derived in Sec.~\ref{sec:jamSAT}, which indeed must emerge from the regular zero temperature UNSAT solution upon
approaching the jamming point.

To summarize, when $\c\to \io$ and in the region of $q \to 1$, we expect two different scaling solutions: 
when $1-q \ll 1-q_*$, we have the ``regular'' UNSAT scaling of Sec.~\ref{sec:scaling_UNSAT}; for $1-q \gg 1-q_*$ we have
instead the ``jamming'' scaling solution of Sec.~\ref{sec:jamSAT}. 
The matching point $q_*$ is determined by the condition that $\chi P(1,0) \sqrt{1-q_*}  \sim 1$, and
the value of $y(q_*)$ is then simply
\beq
y(q_*) = \frac{\c P(1,0)}{\sqrt{1-q}} \sim \frac1{1-q_*} \ .
\eeq
Note that in the jamming solution, Eq.~\eqref{eq_scaling_sol}, we have $P(q_*,0) \sim (1-q_*)^{-a/\k}$, and we know from Eq.~\eqref{eq:dPh} that
in the regular solution $P(q,h)$ changes by a very small amount, $\sim \sqrt{1-q}$, when
$q\to 1$. We conclude that $P(1,0) \sim (1-q_*)^{-a/\k}$ and more
generally $P(1,h)\sim
(1-q_*)^{(1-\kappa)/\kappa}p_-((1-q_*)^{(1-\kappa)/\kappa}h)$ for
negative values of $h$. We obtain therefore
\beq\label{eq:crucial123}
\chi P(1,0) \sqrt{1-q_*}  \sim \chi (1-q_*)^{1/2-a/\k} \sim 1
\hskip20pt
\Rightarrow
\hskip20pt
\chi \sim (1-q_*)^{a/\k-1/2}
\hskip20pt
\Rightarrow
\hskip20pt
1-q^* \sim \c^{\frac{2\k}{2a-\k}}
\ ,
\eeq
which concludes the analysis of the matching between the two scaling solutions on the UNSAT side of the transition.

\subsection{Scaling relations between exponents}
\label{sec:scaling}

The relation $a=1-\k/2$, and its consequences that are given in Eq.~\eqref{eq:ak}, has been found in Ref.~\cite{CKPUZ14} within numerical precision by solving
the equations for the critical exponents derived in Sec.~\ref{sec:Jproof}. 
Eq.~\eqref{eq:ak} is physically very important:
in fact, the relation $\g=1/(2+\theta)$
has been proven in~\cite{Wy12} to be a direct consequence of marginal stability (in the perceptron, the same marginal stability argument
is discussed in~\cite{FP15}).
It would therefore be nice to have a more direct analytical proof of the relation $a=1-\k/2$, that does not rely on a numerical calculation.

While in principle it must be possible to obtain such a proof directly from the properties of the equations of Sec.~\ref{sec:Jproof}, that
define all the critical exponents, here we give an independent argument by showing 
how the matching condition between the two
asymptotic solutions allows one to derive analytically this relation.
To this aim, we focus on the pressure $p = -[h]$. In Appendix~\ref{app:pressure}, we show that $p \propto 1/\chi^2$
in the jamming limit $\chi\to\io$.
 Using now $P(1,h)\approx
 (1-q_*)^{(1-\kappa)/\kappa}p_-((1-q_*)^{(1-\kappa)/\kappa}h)$ for
 $h<0$, which was derived in Sec.~\ref{sec:jammingUNSAT}, together with Eq.~\eqref{gab_distr}, we find
 \begin{eqnarray}
   \label{eq:crux}
   \r(h<0) \sim \chi (1-q_*)^{(1-\kappa)/\kappa}p_-[\chi (1-q_*)^{(1-\kappa)/\kappa}h]
   \qquad 
   \Rightarrow
   \qquad
   [h]\sim \frac {1}{ \chi  (1-q_*)^{(1-\kappa)/\kappa}} \ . 
 \end{eqnarray}
Because we know that $[h]\propto 1/\chi^2$, we must have
\begin{eqnarray}
  \label{eq:7}
\chi\sim   (1-q_*)^{(1-\kappa)/\kappa}\:.
\end{eqnarray}
This is compatible with (\ref{eq:crucial123}) only if $a=1-\k/2$, which
gives an independent proof of Eq.~\eqref{eq:ak}.
Finally, recalling that $P(q_*,h) \sim P(1,h)$, 
and using Eq.~\eqref{eq:7} into Eq.~\eqref{eq_scaling_sol}, we obtain
\beq\label{eq:95}
P(1,h)  \sim P(q_*,h) \sim \begin{cases}
\c \, p_-( h \c ) & \text{for } h \sim -\c^{-1}  \\
\c^{-\frac{2 - \k}{2(1-\k)}} p_0\left( h \c^{-\frac{\k}{2(1-\k)}} \right) & \text{for } |h| \sim \c^{\frac{\k}{2(1-\k)}} \\
p_+(h) & \text{for } h \gg \c^{\frac{\k}{2(1-\k)}}
\end{cases}
\eeq
which holds on the UNSAT side upon approaching jamming.

\subsection{Scaling of several interesting observables at the jamming transition}
\label{sec:gammatheta}

We have now fully characterized the scaling of the basic quantities, $x(q)$, $\l(q)$, $f(q,h)$, and $P(q,h)$, in the vicinity
of the jamming transition, both on the SAT and on the UNSAT side. 
On the SAT side, our main results are Eqs.~\eqref{eq:scalSAT} and \eqref{eq_scaling_sol}, together with
Eqs.~\eqref{eq:73}, \eqref{eq:74} and \eqref{eq:ak}.
On the UNSAT side, our main results are Eqs.~\eqref{eq:9}, \eqref{eq:chiS}, \eqref{gab_distr} and \eqref{eq:95}.
From these results, the scaling of all the interesting observables can be derived.
In this section, we focus on some of the most studied observables in the context of jamming, and we show that
all the known results are reproduced by the scaling solution.

\subsubsection{Energy, pressure, number of contacts}

We start by considering the energy, the pressure and the number of contacts defined in Eq.~\eqref{eq:zpe}.
From Eqs.~\eqref{gab_distr} and \eqref{eq:95}, we deduce that in the jamming limit from the UNSAT (jammed) phase,
\beq
\r(h<0) \sim \chi^2 p_-(\chi^2 h) \qquad
\Rightarrow
\qquad [h^n] \propto \c^{-2n} \ .
\eeq
In particular, the energy and pressure scale as
\beq\label{eq:pejammed}
p = - [h] \sim \c^{-2} \int_{-\io}^0 \de t \, p_-(t) |t| \ ,\qquad  
e = \frac{\a}2[h^2] \sim \frac{\a}2 \c^{-4}\int_{-\io}^0 \de t \, p_-(t) t^2 \propto p^2 \ .
\eeq
Note also that from Eq.~\eqref{eq:chiS} we have for the isostatic index:
\beq\label{eq:ciso}
c = \a z = \a [1] = \left(1 + \frac1\c\right)^2 \sim 1 + \frac{2}\c \ .
\eeq
We deduce that {\it at jamming the system is isostatic}, and that
 the excess of contacts in the jammed phase scales like $c - 1 \propto \c^{-1} \propto p^{1/2}$.
To obtain the complete phenomenology of jamming, one should also prove that the pressure $p \propto \c^{-2}$ 
vanishes linearly in the distance from jamming in the $(\a,\s)$ plane: we did not find a proof of this relation,
which at present must be derived from the numerical solution of the equations. Apart from that, the scaling relations
$e \propto p^2$ and $c-1 \propto p^{1/2}$ perfectly agree with numerical observations in jammed sphere
packings~\cite{OSLN03,LNSW10}.

In the SAT phase, at zero temperature, there are by definition no negative gaps, $\r(h<0)=0$; consequently pressure, energy, and contacts all vanish. However, one can study the limiting values of these
quantities when $T\to 0$. As an example, let us consider the pressure, which is a standard observable
in particle systems.
In the perceptron, it can be written as the derivative of the free energy with respect to $\s$, as
it can be checked directly from the definition of the partition function in Eq.~\eqref{eq:Zdef}. Using the replica expression of the free
energy in Eq.~\eqref{fe}, one then obtains:
\beq
p = -[h] = \frac1\a \frac{\de {\rm f}}{\de \s} 
= - T \frac{\de}{\de \s} \int \de h \g_{q_m}(h) f(0,h-\s) =T \int \de h \g_{q_m}(h) f'(0,h-\s) 
=T \int \de h P(0,h) f'(0,h) 
\ .
\eeq
Using the equations for $P$ and $f$, it is possible to show that 
\beq
0 = \frac{\de}{\de x}  \int \de h P(x,h) f'(x,h)
\qquad
\Rightarrow
\qquad
p = T \int \de h P(1,h) f'(1,h) \ .
\eeq
The proof is obtained by writing the derivative as $\int \de h (\dot P f' + P \dot f')$, using the equations 
for $P$ and $f$ and integrating by parts~\cite{RU16}.
Therefore, in the SAT phase the pressure vanishes proportionally to $T$, a standard result for hard spheres
(which correspond to the $T\to 0$ limit of soft spheres in the SAT phase). 
However approaching the jamming line the ratio $p/T$ (also called ``reduced pressure'' or ``compressibility factor'' in the hard spheres
literature) diverges.
Inserting the scaling form given by
Eq.~\eqref{eq_scaling_sol}, noting that $\l(1) = 1-q_M = \ee^\k$,
one has
\beq
p = T \int \de h P(1,h) \frac{m(1,h)}{\l(1)} = \frac{T}{\ee^\k} \int \de h P(1,h) m(1,h)  \ .
\eeq
The scaling form in Eq.~\eqref{eq_scaling_sol} is cutoff when $1-q \sim \ee^\k$, and therefore the same scaling
holds for $P(1,h)$ and $m(1,h)$ with $1-q \to \ee^\k$. Then, in the $\ee\to 0$ limit the region $h>0$ does not
contribute to the integral because $m(1,h)\to 0$ while $P(1,h)$ stays finite. One can check that the matching region
also gives a subdominant contribution. The leading term is associated to the $h<0$ region, where $m(1,h) \to -h$ and
\beq
p = \frac{T}{\ee^\k} \int_{-\io}^0 \de h \ee^{1-\k} p_-(h \ee^{1-\k}) |h| = \frac{T}{\ee} \int_{-\io}^0 \de t \, p_-(t) |t| \ .
\eeq
We conclude that the reduced pressure $p/T$ diverges proportionally to $\ee^{-1}$, and therefore $1-q_M \propto (p/T)^{-\k}$,
a result that has been numerically tested in hard sphere systems~\cite{CKPUZ14}. This provides a physical meaning
for the exponent $\k$.
In a similar way one can study
the scaling of the energy and the number of contacts in the SAT phase.

\subsubsection{Force and gap distributions}

\begin{figure}
\centering
\includegraphics[scale=1.2]{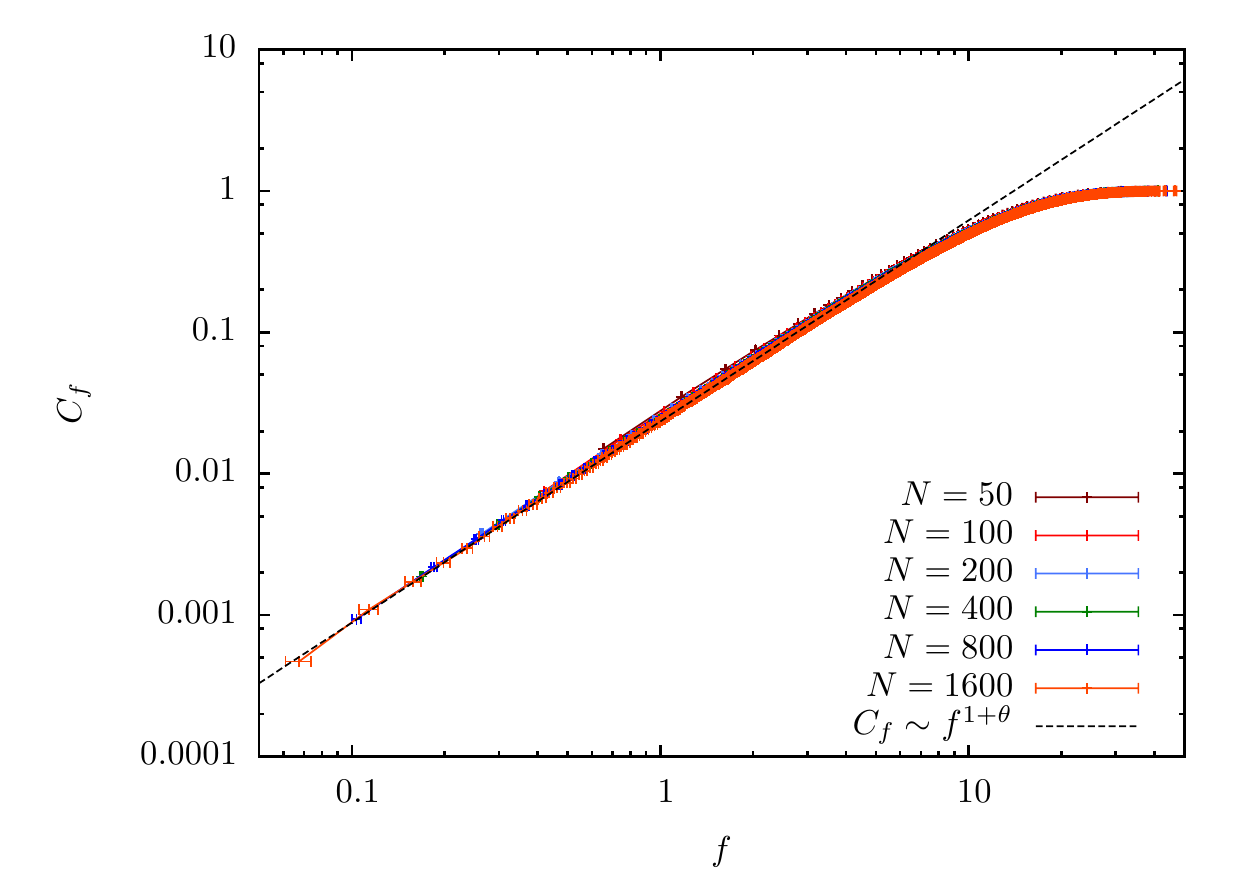}
\caption{Plot of the cumulative of the force distribution $C_f=\int_0^f \de \tilde f p(\tilde f)$ for $N=50,\dots 1600$.
The line is proportional to the theoretical prediction
$C_f\sim f^{1+\theta}$.}
\label{plot_forces}
\end{figure}

In the UNSAT (jammed) phase, positive gaps correspond to satisfied constraints, while negative gaps can be associated to contact forces
according to Eq.~\eqref{eq:fmudef}. The distribution of small positive gaps and small contact forces has been associated with important
properties of the packings, including marginal stability~\cite{Wy12,LDW13}. In this section we discuss these distributions. They can be 
straightforwardly derived from Eq.~\eqref{gab_distr} and Eq.~\eqref{eq:95} which together imply, for $\chi \gg 1$:
\beq
\r(h) \sim \begin{cases}
\c^2 p_-(\c^2 h ) & \text{for } h< 0 \ , \\
p_+(h) & \text{for } h > 0 \ .
\end{cases}
\eeq
At jamming, when $\c\to\io$, the distribution of gaps concentrates on $h\geq 0$, where it is given by
\beq
\r(h) = [1] \d(h) + p_+(h) \ ,
\qquad
h\geq 0 \ ,
\eeq
where $[1] = \int_{-\io}^0 \de h \r(h) = \int_{-\io}^0 \de t p_-(t) = 1/\a$ according to Eq.~\eqref{eq:ciso}. 
Isostaticity is expressed by the presence of a delta peak in $\r(h)$ whose weight is the
fraction of contacts. Moreover, this implies that the distribution of small positive (satisfied) gaps, according to Eq.~\eqref{eq:73}, behaves as a power-law $\r(h\to 0^+) \sim h^{-\g}$
with $\g = (2-\k)/\k = 0.41269\ldots$.
Similarly, the normalized probability distribution of scaled contact forces $f^s_\mu = -[1] h_\mu/p$, with $p/[1]$ given by Eq.~\eqref{eq:pejammed}, 
can be deduced from $\r(h<0)$ and reads
\beq
p(f) = \a \bar f \, p_-\left(-f \bar f \right) \ ,
\qquad 
\bar f = \a\int_0^\io \de t \, p_-(-t) \, t \ ,
\qquad 
\frac1{\a} = [1] = \int_0^\io \de t \, p_-(-t) \ .
\eeq
Note that $\int_0^\io \de f p(f) =\int_{-\io}^0 \de f p(f) \, f = 1$. This implies, according to Eq.~\eqref{eq:74}, that the distribution of small forces
behaves as a power-law, $p(f\to 0^+) \sim f^\th$, with $\th = 1/\g-2 = 0.42311\ldots$.
These results give a physical interpretation to the critical exponents $\g$ and $\th$, and connect with the results of~\cite{Wy12,LDW13,CCPZ12,CCPZ15,Ka16}.
Similar results can be obtained upon approaching the jamming transition from the SAT phase, with the only difference that in that case one has to
define the forces by separating the positive gaps into those who become contacts at jamming and those who stay positive~\cite{BW06,CKPUZ13,CCPZ12}.
{As a check of the theoretical predictions we have extended the
numerical simulations of \cite{FP15} to large systems: in Fig.~\ref{plot_forces} we compare the cumulative distribution $C_f=\int_0^f \de \tilde f p(\tilde f)$
for sizes from $N=50$ to
$N=1600$ with the theoretical prediction.}

We conclude by mentioning that one can also study the vibrational spectrum, both in the SAT and UNSAT phases~\cite{FPUZ15,AFP16}, to discuss
the presence of soft modes associated to both fullRSB and jamming, and connect the density of states with these critical exponents as it is also observed
in sphere packings~\cite{DLBW14}. Also, it has been shown in~\cite{FS16} that the asymptotic behavior of the function $y(q)$, related to the exponent
$\k$, controls the scaling of the avalanches at the jamming point, which is therefore different than the scaling in the UNSAT phase.

\section{Conclusions}

In this work we have formulated the jamming problem as a general constraint satisfaction problem with continuous variables (Sec.~\ref{sec:CCSP}).
We have then specialized on the random perceptron, a well known machine learning model, which is a prototype of this class of problems (Sec.~\ref{sec:perc_def}).
In the non-convex regime, the model shows a complex zero temperature phase diagram in the plane of the two control parameters $(\a,\s)$,
which has been fully characterized in Sec.~\ref{sec:PD}.
In particular, we have shown that for $\s<0$ and large enough $|\s|$, the phase diagram as a function of $\a$ shows the characteristic phenomenology
associated with the Random First Order Transition (RFOT) mean field theory of glasses~\cite{BB11,Ca09,PZ10,WL12,KT15}.
Our main result is that
the jamming transition, which can be seen as the SAT/UNSAT threshold, 
 is always associated with full replica symmetry breaking in the non-convex regime.
 In Sec.~\ref{sec:SATUNSAT},
we have thus discussed the scaling behavior of the fullRSB equations around the jamming transition.
We have shown that approaching jamming from the SAT phase,
one obtains the same critical exponents of the jamming transition of hard spheres in high dimension, thus reproducing the results of~\cite{CKPUZ14}. 
Furthermore we have extended the study of~\cite{CKPUZ14} by analyzing 
the model in the UNSAT phase, where we have obtained the scaling solution of the fullRSB equations,
showing that it reproduces the critical behavior of soft spheres, when approaching the transition from the jammed phase~\cite{OSLN03}.
We have provided a complete matching between the scaling solutions in the SAT and UNSAT phases, and derived scaling relations between the critical exponents,
also showing their physical interpretation as the exponents that control the gap and force distributions at jamming~\cite{Wy12}.
Our results are also consistent with the scaling analysis of~\cite{GLS16}. 
These results, together with the results on the vibrational spectrum obtained
in~\cite{FPUZ15,AFP16}, and the study of avalanches performed in~\cite{FS16},
provide a complete study of all the properties of the random perceptron that are relevant for the study of the glass and jamming transition
at the mean field level.

This work opens the way to the study of the jamming transition in other ensembles of random constraint satisfaction problems with continuous variables. 
{The outcome of this study is that the non-convex jamming transition lies always in a fullRSB phase and we conjecture that this happens in a large class of CCSP. Although we are not able to prove it we note that this is what happens not only in the model we have analyzed but also in Hard-Spheres in high dimension \cite{CKPUZ14}.}
Another natural question that arises is to what extent the scaling behavior that we have found is universal.
The fact that the critical exponents at the jamming transition for the random perceptron and hard spheres coincide support the conjecture that 
in non-convex random CSPs,
the jamming point is highly universal.
This conjecture has been tested and confirmed in generalized
perceptron models with multibody interaction~\cite{HaA}.
There could be, however, different universality classes: understanding which features of the models determine them is a very important
direction for future work.
Going beyond the mean field, infinite dimensional level, for example by considering random dilute versions of the
perceptron, is another interesting direction for future research.

\acknowledgments

This work was supported by grants from
the Simons Foundation (No. 454941, S. F.; No. 454949, G. P.; No. 454955, F. Z.).
Ce travail a b\'en\'efici\'e d'une aide Investissements d'Avenir du LabEx PALM (ANR-10-LABX-0039-PALM) (P. Urbani).


\clearpage
\appendix

\section{Derivation of the replicated free energy} 
\label{sec:App1}

In this Appendix, we give a detailed derivation of the replica equations for the perceptron model.
Starting from the partition function in Eq.~\eqref{eq:Zdef}, 
and
neglecting proportionality constants, the replicated partition function can be written as
\beq
\overline{Z^n} \propto  \int \left(\prod_{a} \DD\vec X_a \right) \left(\prod_{a,\m} dr_a^\m d\hat r_a^\m \right)
\overline{ e^{\sum_{a\m} i \hat r_a^\m ( r_a^\m - N^{-1/2}  \vec X_a \cdot \vec\x^\m )} } \prod_{a\m} e^{-\b v(r_a^\m - \s)} \ ,
\eeq
where $a=1\cdots n$ and $\m = 1\cdots M$; one can check that integrating over $\hat r_a^\m$ produces delta
functions that fix $r_a^\m$ as in the original partition function.
Next, the Gaussian integral over the quenched disorder $\vec \x^\m$ gives
\beq
\overline{ e^{- N^{-1/2}  \sum_{a\m} i \hat r_a^\m   \vec X_a \cdot \vec\x^\m } } = e^{-\frac12 \sum_{ab\m} \hat r_a^\m \hat r_b^\m Q_{ab}} \ ,
\hskip20pt
Q_{ab} = \frac1N \vec X_a \cdot \vec X_b \ .
\eeq
We obtain then
\beq\begin{split}
\overline{Z^n} &\propto  \int \left(\prod_{a} \DD\vec X_a \right)
\left[ 
\int \left(\prod_{a} dr_a d\hat r_a \right)
e^{\sum_{a} i \hat r_a r_a -\frac12 \sum_{ab} \hat r_a \hat r_b Q_{ab} -\b \sum_{a} v(r_a - \s)}
\right]^M \\
&\propto \int dQ_{ab} e^{\frac{N}2 \log\det Q} \left[ 
\int \left(\prod_{a} dr_a d\hat r_a \right)
e^{\sum_{a} i \hat r_a  r_a -\frac12 \sum_{ab} \hat r_a \hat r_b Q_{ab} -\b \sum_{a} v(r_a - \s)}
\right]^M \ ,
\end{split}\eeq
where we made a change of variables from $\vec X_a$ to $Q_{ab}$, with Jacobian $\exp(\frac{N}2 \log\det  Q)$.
Finally, one can easily show by developing both sides in powers of $Q_{ab}$ that
\beq
e^{ -\frac12 \sum_{ab} \hat r_a \hat r_b Q_{ab} } = \left. e^{ 
\frac12 \sum_{ab} Q_{ab} \frac{\partial^2}{\partial k_a \partial k_b}
}
e^{ - \sum_a k_a i \hat r_a}
\right|_{k_a=0} \ ,
\eeq
and therefore
\beq\begin{split}
\int \left(\prod_{a} dr_a d\hat r_a \right) &
e^{\sum_{a} i \hat r_a  r_a -\frac12 \sum_{ab} \hat r_a \hat r_b Q_{ab} -\b \sum_{a} v(r_a - \s)}
 \\
& =  \int \left(\prod_{a} dr_a d\hat r_a \right) 
\left. e^{ 
\frac12 \sum_{ab} Q_{ab} \frac{\partial^2}{\partial k_a \partial k_b}
}
e^{ \sum_a (r_a - k_a) i \hat r_a -\b \sum_{a} v(r_a - \s)}
\right|_{k_a=0} \\
& \propto  
\left. e^{ 
\frac12 \sum_{ab} Q_{ab} \frac{\partial^2}{\partial k_a \partial k_b}
}
e^{  -\b \sum_{a} v(k_a - \s)}
\right|_{k_a=0} \ ,
\end{split}\eeq
where to obtain the last line we integrated over $\hat r_a$ to obtain a $\d(r_a - k_a)$ and then
integrated over $r_a$. 
Introducing $h_a = k_a - \s$ and setting $\a = M/N$, we obtain the final result for the replicated partition function
\beq\begin{split}
& \overline{Z^n} = \int dQ e^{N S(Q)} \ , \\
& S(Q) = \frac12 \log\det Q + \a \log \left( \left.
e^{\frac12 \sum_{ab} Q_{ab} \frac{\partial^2}{\partial h_a \partial h_b}}
\prod_a e^{-\b v(h_a)}
\right|_{h_a=-\s} \right) \ ,
\end{split}\eeq
that is reported in Eq.~\eqref{eq:SQdef}.

\section{Numerical solution of the equations} \label{sec:KRSB}

\begin{figure}[t]
\includegraphics[width=.4\textwidth]{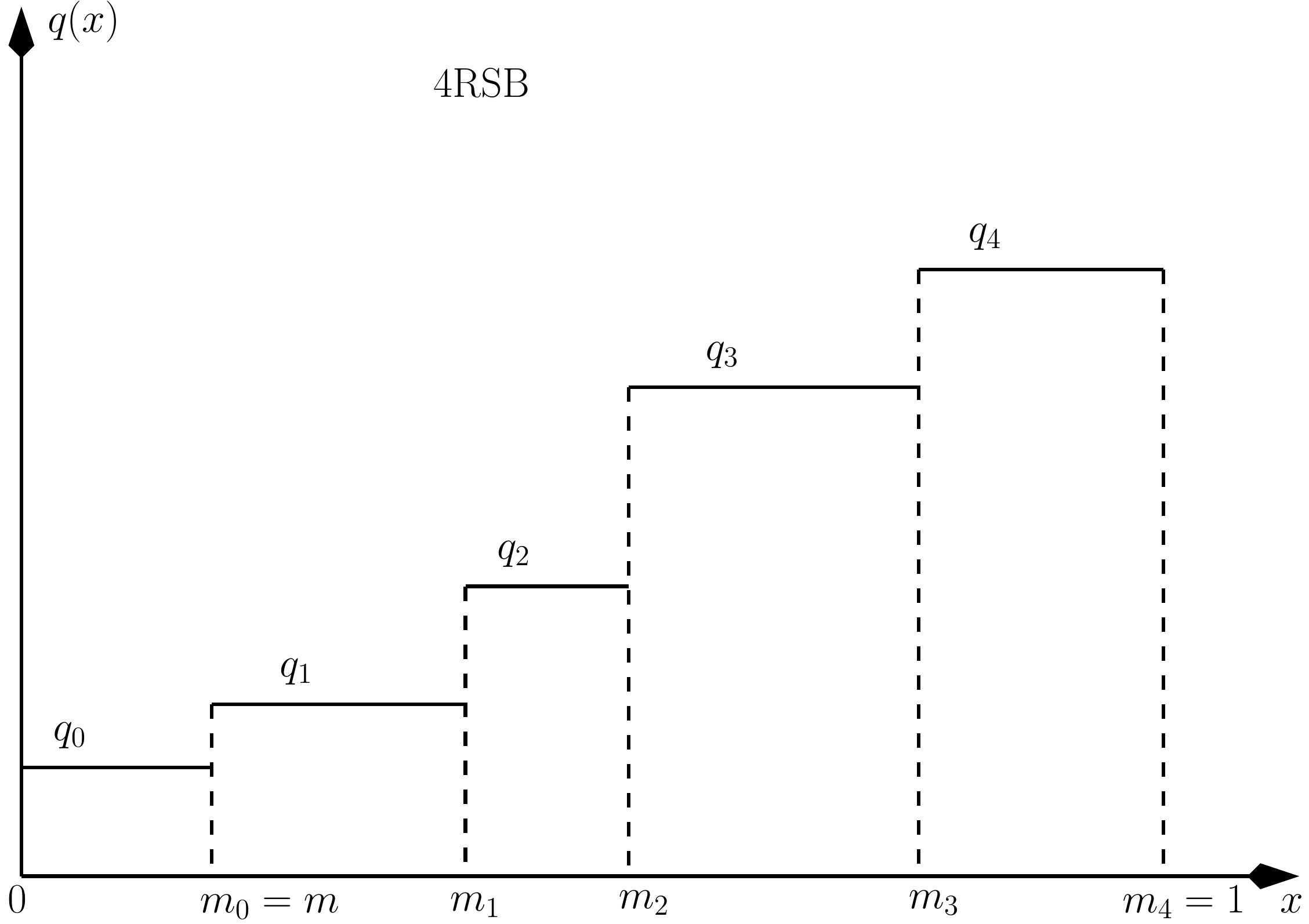}
\caption{An example of the parametrization of the matrix $Q_{ab}$ for a $4$RSB case.
The reader should keep in mind that we use the convention $m_0\equiv m$, $m_{k+1} \equiv 1$.
}
\label{fig:hierD}
\end{figure}
 
 In this Appendix, we derive the equations corresponding to a discrete $K$RSB ansatz, 
illustrated in Fig.~\ref{fig:hierD}. This
 is particularly useful for the
numerical solution.

For a function $q(x)$ with $K$RSB form, the free energy of Eq.~\eqref{fe} is
\beq
s[q(x)] = \lim_{n\to 0} \frac1n S(Q) = \frac12 \left[
\log(1 - \la q \ra) + \frac{q_m}{1 - \la q \ra} + \int_0^1 \frac{\de y}{y^2} \log\left(
\frac{ \l(y) }{1 - \la q\ra}
\right)
\right] + \a f_S(-\s)
\eeq
with 
$\la q\ra = \int_0^1 \de x \, q(x)$, and $\l(x)$ is also a piecewise constant function whose values
can be derived from $q(x)$.
The function $f(m_i,h)$ is the discrete version of $f(q,h)$ and it satisfies the discrete version
of Eq.~\eqref{eqf}:
\beq\label{eq:f}
\begin{cases}
f(1,h) = \log \g_{1 - q_k} \star e^{-\b v(h)} \ , \\
f(m_i,h)  = \frac{1}{m_i} \log \left[ \g_{q_{i+1} - q_{i}} \star e^{m_i f(m_{i+1},h)} \right] \ , \hskip20pt i=k-1,\cdots,0 \ , \\
f_S(h) = \g_{q_0} \star f(m_0,h) \ .
\end{cases}
\eeq
The saddle point equation is
\beq
Q^{-1}_{cd} = - \a \frac{ \left( \left.
e^{\frac12 \sum_{ab} Q_{ab} \frac{\partial^2}{\partial h_a \partial h_b}}
\frac{\partial^2}{\partial h_c \partial h_d} \prod_a e^{-\b v(h_a)}
\right|_{h_a=-\s} \right) }
{
\left( \left.
e^{\frac12 \sum_{ab} Q_{ab} \frac{\partial^2}{\partial h_a \partial h_b}}
\prod_a e^{-\b v(h_a)}
\right|_{h_a=-\s} \right)
} \equiv -\a M_{cd} \ ,
\eeq
and
the matrix $M_{ab}$ is also a hierarchical matrix whose components can be written as
\beq
M_i = \int \de h P(m_i,h) f'(m_i,h)^2 \ ,
\eeq
with
\beq\label{eq:P}
\begin{cases}
P(m_0,h)  = \g_{q_0}(h+\s) \\
P(m_i,h) = e^{ m_{i-1} f(m_i,h) }  \g_{q_{i}-q_{i-1}} \star \left[ P(m_{i-1},h) e^{-m_{i-1} f(m_{i-1},h)} \right] \ ,
\hskip20pt i=1,\cdots,k \ ,
\end{cases}
\eeq
which provides a discrete version of Eq.~\eqref{eq:qP}.
Note that all the $P(m_i,h)$ are normalized to 1.
The saddle point equation for a hierarchical matrix are therefore
\beq\label{qm1}
q^{-1}_i = -\a M_i = -\a \int \de h P(m_i,h) f'(m_i,h)^2 \ .
\eeq
To close the equations, we now need to express $q_i$ as a function of $q^{-1}_i$. 

We start from the exact relations (here $q_d=1$ is the diagonal element and $[q](x) = x q(x) - \int_0^x \de y q(y)$):
\beq\label{eq:ldef}
\begin{split}
\l(x) & \equiv q_d - xq(x) - \int_x^1 \de y q(y) \hskip15pt \Leftrightarrow \hskip15pt
q(x) = q_d - \frac{\l(x)}x + \int_x^1 \frac{\de y}{y^2} \l(y) \ ,
\\
(q^{-1})_d &= \frac{1}{\l(0)} \left( 1 - \int_0^1  \frac{\de y}{y^2}  \frac{[q](y)}{\l(y)} - \frac{q(0)}{\l(0)} \right) \ , \\
(q^{-1})(x)&=-\frac{1}{\l(0)}\left[\frac{q(0)}{\l(0)}+\frac{[q](x)}{x \, \l(x)}+
\int_0^x\frac{\de y}{y^2}\frac{[q](y)}{\l(y)}\right] 
= - \frac{q(0)}{\l(0)^2} - \int_0^x \frac{\de y}{\l(y)^2} \dot q(y)
\ . 
\end{split}\eeq
The above equalities are obtained by showing that the derivatives with respect to $x$, as well as the values
in $x=0$ or $x=1$ coincide.
From these one can derive several useful identities:
\beq
(q^{-1})_d - \la q^{-1} \ra - [q^{-1}](x) = \frac{1}{q_d -\langle q\rangle-[q](x)}  \hskip10pt \Rightarrow \hskip10pt
(q^{-1})_d - \la q^{-1} \ra  = \frac{1}{q_d -\langle q\rangle}   \ . 
\eeq
From the last Eq.~\eqref{eq:ldef} we get $(q^{-1})(0) = - \frac{q(0)}{\l(0)^2}$. Collecting these two relations we get
\beq
\l(0) = \sqrt{- \frac{q(0)}{(q^{-1})(0)} } \ , \hskip30pt 
\frac{1}{\l(x)} = \frac{1}{\l(0)} - [q^{-1}](x) \ .
\eeq
 This relations allows one to reconstruct
$\l(x)$ from $(q^{-1})(x)$ and using the first Eq.~\eqref{eq:ldef} we can obtain $q(x)$ from $\l(x)$.
In the discrete case one has $[q]_i = \sum_{j=0}^{i-1} m_j (q_{j+1} - q_j )$ and then
these equations become
\beq\label{eq:[q]i}
\begin{split}
\l_0 &= \sqrt{- \frac{q_0}{q^{-1}_0} } \ , \hskip30pt 
\frac{1}{\l_i} = \frac{1}{\l_{i-1}} - m_{i-1} (q^{-1}_i - q^{-1}_{i-1}) \ . \\
q_i &= 1 - \frac{\l_i}{m_i} - \sum_{j=i+1}^k \left(\frac1{m_j} - \frac1{m_{j-1}}\right) \l_j \ .
\end{split}\eeq
The procedure to solve these equations is therefore (for a fixed grid of $m_i$):
{\it (i)} start from a guess for $q_i$, 
{\it (ii)} solve Eqs.~\eqref{eq:f} and \eqref{eq:P} to obtain $f(m_i,h)$ and $P(m_i,h)$,
{\it (iii)} from Eq.~\eqref{qm1} compute $q^{-1}_i$, and {\it (iv)} use Eqs.~\eqref{eq:[q]i} to compute the new $q_i$.

\section{Asymptotic behavior in the UNSAT phase} 
\label{App_Asy}

In this Appendix we discuss the asymptotic behavior of $f(q_M,h)$ in the UNSAT phase where $q_M=1-\chi T$.
This is given by the zero temperature limit of
\beq
f(q_M,h)=\log \int_{-\infty}^\infty \frac{\de z}{\sqrt{2\pi \chi T}}\exp\left[-\frac{(h-z)^2}{2\chi T} -\frac{z^2}{2T }\th(-z)\right] \ .
\eeq
For $T\to 0$ and $h\sim \OO(1)$ we can compute the integral using a saddle point approximation.
The saddle point equation is
\beq
\frac{z-h}{\chi} +z \th(-z)=0 \ .
\eeq
Its solutions are given by
\beq
\begin{split}
z^*(h)&=h \ , \ \ \ \ \ \ \ \ \ \ \  \ \ \ \ \mathrm{for} \ \ h>0 \ ,\\
z^*(h)&=\frac{h}{1+\chi} \ , \ \ \ \ \ \ \ \ \ \mathrm{for} \ \ h<0 \ .
\end{split}
\eeq
Thus, defining $t= z-z^*(h)$ obtain
\beq
\begin{split}
f(q_M,h)&=\log \left[ \th(h)\int_{-\infty}^{\infty} \frac{\de t}{\sqrt{2\pi \chi T}}e^{-t^2/(2\chi T)} + \th(-h)e^{-h^2/(2(1+\chi)T)}\int_{-\infty}^{\infty} \frac{\de t}{\sqrt{2\pi \chi T}}e^{-\frac{1+\chi}{2\chi T}t^2} \right]\\
&\simeq \left(-\frac{h^2}{2(1+\chi)T}-\frac 12 \ln(1+\chi) \right)\th(-h) \ .
\end{split}
\eeq

\section{Scaling of the pressure in the UNSAT phase}
\label{app:pressure}

In this Appendix, we prove that the pressure satisfies the relation $p = -[h] \propto 1/\chi^2$,
a relation used to prove the scaling relation $a=1-\k/2$ in Sec.~\ref{sec:scaling}.
To this aim, 
let us modify the problem by replacing the hard spherical constraint by a Lagrange multiplier $\mu$, and
consider the following exact relation
\beq
0=\frac{1}{Z N}\int \de \vec X \sum_{i=1}^N \frac{\partial}{\partial X_i} \left\{ X_i \exp\left[-\b \left(H[\vec X]-\frac{\mu }{2}\sum_{j=1}^N (X^2_i -1)\right)\right]\right\} \ ,
\label{exact_mu}
\eeq
from which it follows that
\beq
\mu=-T +[h^2]+\s [h] \ .
\eeq
We will prove below that
  in the zero temperature limit and in the fullRSB phase, $\mu=1/\chi^2$,
so that the following exact zero temperature relation holds:
\beq
1=\chi^2\left([h^2]+\s [h]\right)\:.
\eeq
Because
close to jamming $[h^2]\ll [h]$ this reduces to $\lim_{\chi\to\io}\chi^2 \sigma
 [h]=1$ at jamming, which proves $[h]\propto 1/\chi^2$.

We now compute $\mu$ using the replica method.
Because $X^2$ is not constrained, we have that $Q_{aa}=q_d$ and we need to find an additional variational equation for $q_d$.
Let us write down the free energy as a function of $q_d$ and $\mu$. We have
\beq
\begin{split}
s[q_d, q(x)] &= \frac 12\left[\log\left(q_d-q_M\right) + \frac{q_m}{q_d-\langle q \rangle}+\int_0^1 \de x \frac{\dot q(x)}{\l(x)}\right]
+\alpha \gamma_{q_m}\star f(0,h) |_{h=-\sigma} +\frac 12\b \mu(q_d-1)\\
&
-\alpha \int \de h\; P(q_M,h)\;[f(q_M,h) - \log \g_{q_d - q_M} \star e^{-\b
  v(h)}]\\
&
+\alpha  \int \de h\;\int_{q_m}^{q_M} \de q\;  P(q,h)\;[\dot f(q,h)+\frac 12 \left[ f''(q, h) + x(q) f'(q, h)^2    \right]  ].
\end{split}
\eeq
where now
\beq
\begin{split} \label{lambdatilde}
\l(x)&=q_d- x q(x)-\int_x^1\de y q(y)\\
\l(q) &=  q_d - q_M + \int_q^{q_M} \de p x(p) \ .
\end{split}
\eeq
The variational equation for $q(x)$, $f(q,h)$ and $P(q,h)$ do not change except for the initial condition for $f$ which is
\beq
f(q_M,h)=\log \g_{q_d - q_M} \star e^{-\b
  v(h)}\
\eeq
and $\l(q)$ is given by Eq.~(\ref{lambdatilde}).
Note that the variational equation with respect to $\mu$ fixes the spherical constraint $q_d=1$.
At this point we can take the variation with respect to $q_d$ to get the following equation
\beq
0=\frac 12 \left[\frac{1}{q_d -q_M}-\frac{q_m}{\left(q_d-\langle q\rangle\right)^2}-\int_0^1\de u \frac{\dot q(u)}{\l^2(u)}\right]+\a\int \de h P(q_M,h)\frac{\partial}{\partial q_d}\log \g_{q_d - q_M} \star e^{-\b v(h)}+\frac{\b \mu}{2} \ .
\eeq
It is very easy to show that the last term of the equation above can be rewritten as
\beq
\a\int \de h P(q_M,h)\frac{\partial}{\partial q_d}\log \g_{q_d - q_M} \star e^{-\b v(h)}=\frac \a 2 \int \de h P(q_M,h)\left[m'(q_M,h)+m^2(q_M,h)\right]
\eeq
where again $m(q_M,h)=\partial f(q_M,h)/\partial h$.
At this point we can use the saddle point equations (\ref{eq:qx1}) and $\eqref{eq:qx2}$ to write
\beq
\begin{split}
\frac{q_m}{\l^2(q_m)}+\int_{q_m}^{q_M}\frac{\de p}{\l^2(p)}&=\a \int \de h P(q_M,h)m^2(q_M,h) \ ,\\
\frac{1}{(q_d -q_M)^2}&=\a\int \de h P(q_M,h)\left(m'(q_M,h)\right)^2 \ .
\end{split}
\eeq 
Using that the variational equation over $\mu$ gives $q_d=1$, that $q_M=1-\chi T$ and that
\beq
m'(q_M,h)=-\frac{\b }{1+\chi}\th(-h)
\eeq
we get $\mu=1/\chi^2$.

\end{document}